\documentclass[preprint,12pt]{myclass}

\usepackage{lineno}
\usepackage{graphicx,xcolor}
\usepackage{amssymb,latexsym}
\usepackage{mathrsfs}
\usepackage{soul}
\usepackage{natbib}
\usepackage{amsbsy}
\usepackage{psfrag}
\usepackage{mathtools}
\usepackage{color}
\usepackage{amsmath}
\usepackage[version=4]{mhchem}
\usepackage{siunitx}
\usepackage{subcaption}
\usepackage{comment}
\usepackage{tikz}
\usepackage{amsthm}
\usepackage{xr}

\usetikzlibrary{calc,arrows.meta,shapes}
\tikzset{ >={Latex[length=.2cm]} }

\definecolor{myC0}{HTML}{003F5C}
\definecolor{myC1}{HTML}{BC5090}
\definecolor{myC2}{HTML}{FFA600}

\providecommand{\sQmm}{\hat{Q}^{-}}
\providecommand{\sQm}{\hat{Q}}
\providecommand{\Qm}{\boldsymbol{\hat{Q}}}
\providecommand{\Qmm}{\boldsymbol{\hat{Q}}^{-}}
\providecommand{\Qmml}{\boldsymbol{\hat{Q}}^{-l}}
\providecommand{\Qmmp}{\boldsymbol{\hat{Q}}^{-p}}
\providecommand{\qmm}{\boldsymbol{\hat{q}}^{-}}
\providecommand{\Qmmr}{\boldsymbol{\hat{Q}}^{-r}}

\providecommand{\qmmp}{\boldsymbol{\hat{q}}^{-p}}
\providecommand{\qmmr}{\boldsymbol{\hat{q}}^{-r}}
\providecommand{\Qf}{\boldsymbol{Q}}
\providecommand{\Qfm}{\boldsymbol{Q}^{-}}
\providecommand{\Em}{\hat E}
\providecommand{\fm}{\hat f}

\providecommand{\dd}{\mathrm{d}}
\providecommand{\um}{\boldsymbol{\breve{u}}}

\providecommand{\bu}{\boldsymbol{u}}

\providecommand{\tw}{0.6}

\begin{document}

\begin{frontmatter}

\title{Limits to extreme event forecasting in chaotic systems}

\author{Yuan Yuan \& Adri\'an Lozano-Dur\'{a}n\\
Department of Aeronautics and Astronautics, Massachusetts Institute of Technology, Cambridge, MA 02139}

\begin{abstract}
%
Predicting extreme events in chaotic systems, characterized by rare
but intensely fluctuating properties, is of great importance due to
their impact on the performance and reliability of a wide range of
systems. Some examples include weather forecasting, traffic
management, power grid operations, and financial market analysis, to
name a few. Methods of increasing sophistication have been developed
to forecast events in these systems. However, the boundaries that
define the maximum accuracy of forecasting tools are still largely
unexplored from a theoretical standpoint. Here, we address the
question: What is the minimum possible error in the prediction of
extreme events in complex, chaotic systems?  We derive the minimum
probability of error in extreme event forecasting
\textcolor{black}{along with its information-theoretic lower and upper
  bounds.} These bounds are universal for a given problem, in that
they hold regardless of the modeling approach for extreme event
prediction: from traditional linear regressions to sophisticated
neural network models.  \textcolor{black}{ The limits in predictability
  are obtained from the cost-sensitive Fano's and Hellman's
  inequalities using the Rényi entropy. The results are also connected
  to Takens' embedding theorem using the \emph{information can't hurt}
  inequality.  Finally, the probability of error for a forecasting
  model is decomposed into three sources: uncertainty in the initial
  conditions, hidden variables, and suboptimal modeling assumptions.} The latter allows us to assess whether \textcolor{black}{prediction
  models} are operating near their maximum theoretical performance or
if further improvements are possible. \textcolor{black}{The bounds are
  applied to the prediction of extreme events in the R\"ossler system
  and the Kolmogorov flow.}
%

%
%
\end{abstract}


\end{frontmatter}


\section{Introduction}

Extreme events, characterized by rare but intensely fluctuating
properties, are ubiquitous in both engineering system and natural
phenomena~\cite{Albeverio2006}. For instance, turbulent gusts over an
aircraft can result in bumpy flights~\cite{Etkin1981}, severe weather
can disrupt communication systems~\cite{Rahmstorf2011}, rare but large
cascades in electrical power grids may lead to
failures~\cite{Schaefer2018}, \textcolor{black}{extreme ocean
  temperature oscillations could impact agriculture and
  ecosystems~\cite{ray2020understanding}, rare but significant
  fluctuations in brain network could cause
  seizures~\cite{ray2022extreme},} and sudden increases in traffic
flow can trigger network paralysis~\cite{Kumar2020}. In these
scenarios, the real-time prediction of extreme events is crucial for
enabling proactive measures to avert potential
issues~\cite{Farazmand2019,Sapsis2021}. By accurately forecasting the
extreme states of dynamical systems, we can mitigate adverse effects,
reduce downtime, and prevent failures.  In this study, we investigate
the limits of predictability in extreme event detection using the
framework of information theory.  The limit obtained is a fundamental
property -- independent of the modeling approach -- that arises from the
finite amount of information the observed state contains about the
extreme event. 

A variety of methods have been employed to predict extreme events in
time series of chaotic dynamical systems. Some of the approaches that
have proven effective include nonlinear dynamics estimation based on
the Koopman operator theory~\cite{Otto2021} and Takens embedding
theorem~\cite{Asch2022}, along with machine learning techniques, such
as support vector machines~\cite{Mukherjee1997}, singular spectrum
analysis and the maximum entropy method~\cite{Ghil2002}.  Advanced
deep learning methods, including auto-encoders~\cite{Racah2017}, long
short-term memory networks~\cite{Vlachas2018}, and reservoir
computing~\cite{Pammi2023} have also been instrumental to devise
forecasting models for chaotic systems with high-dimensional
attractors. A discussion on the role of information in the context of
model prediction and control for chaotic dynamical systems can be
found in Ref.~\cite{LozanoDuran2022}.

Despite the significant advancements described above, the inherent
nature of chaos continues to impose limits on the accuracy of models
for extreme event forecasting. The error of prediction in chaotic
dynamical systems stem from three primary
sources~\cite{Smith2001}. First, the model might not accurately
represent the physical reality. Second, the observable variables may
not capture all the relevant degrees of freedom present in the
dynamical system. Third, the initial conditions required for
forecasting might not be precisely known.

Improvements in the prediction of extreme events can be achieved
either by enhancing models to better represent the physics, gaining
access to more observables, or reducing uncertainty in the initial
conditions. Eliminating modeling errors is theoretically possible,
given the knowledge of a set of governing equations that reflect the
underlying dynamics of the observed system.  However, accessing
variables beyond what is currently observable may be limited by
experimental or computational constraints.  Additionally, no feasible
approach can completely eliminate prediction errors caused by
uncertainty in the initial conditions. Even with highly precise
measurements, minor errors in the initial state eventually amplify due
to chaos, compromising the accuracy of the forecast for long
times~\citep{Lorenz1963}. Here, our focus is not on developing
superior models for extreme event prediction. Instead, we pose the
fundamental question: what is the theoretically maximum achievable
accuracy in extreme event prediction regardless of the modeling
approach and source of error?

\section{Formulation}

\subsection{\textcolor{black}{Modeled extreme event indicator}}

\textcolor{black}{Consider a chaotic dynamical system completely
  determined by $N$ time-dependent variables given by the vector
  $\Qf(t) = [Q_1(t), Q_2(t),\dots, Q_N(t)]$, where $t$ is the time. We
  are interested in the extreme values of the variable $Q_E(t)$, which
  is a function of $\Qf(t)$.} The extreme event indicator $E(t)$ is
defined as
\begin{equation}
E(t)= \begin{cases}1 & \text { if } Q_E(t)\geqslant \eta,
  \\ 0 & \text {otherwise},\end{cases}
\end{equation}
where $\eta$ is the threshold for extreme event
detection. \textcolor{black}{The specific value of the threshold $\eta$
  is dependent on the problem and could be selected based on the
  definition of extreme event for each particular application.}  The
vector of observable variables is defined as \textcolor{black}{$\Qm(t)
  = [\sQm_1(t), \sQm_1(t),\dots, \sQm_M(t)]$}, which contains the
accessible information about the system \textcolor{black}{(i.e., the
  variables that can be measured or are assumed to be
  known)}. \textcolor{black}{The components of $\Qm$ correspond to
  individual components of $\Qf$ or functions of them. In general,
  $M\leq N$ and the number of observed variables $M$ is equal or
  smaller than the number of degrees of freedom of the system $N$}.

We aim to build a predictive model for $E$. To that end, we define the
\textcolor{black}{limited-precision observable} containing information
from the present time and \textcolor{black}{$p\geq 0$ times in the
  past}:
\begin{equation}
\textcolor{black}{\Qmm = [ \Qm(t), \Qm(t-\delta t_1),
  \dots,\Qm(t-\delta t_p)] \pm \delta \Qmm,}
\end{equation}
\textcolor{black}{where $\delta t_i>0, i= 1,\dots,p$ are the time lags
  used for prediction, and $\delta \Qmm$ is the uncertainty in the
  observations. The latter may arise experimentally from inaccuracies
  in measurement tools, numerically from round-off errors in $\Qm$ or
  its discretization, and generally, from any uncertainties in the
  value of $\Qm$}. One could forecast the extreme event indicator in
the future $E(t+\delta t)$ after a time horizon $\delta t>0$ using
$\Qmm$ as the input to the model $\fm$ such that
\begin{equation}
  \Em(t+\delta t) = \fm\left( \Qmm \right),
\end{equation}
where $\Em$ is the modeled extreme event indicator, which might differ
from $E$.  The performance of the model can be evaluated using the
probability of error
\begin{equation}
P_e(\Qmm, \fm) = \text{Probability}( \Em \neq E ) = P(\fm(\Qmm) \neq E).
\end{equation}

Mispredicted extreme events can manifest as either false positives,
$P( \Em = 1, E =0)$, or false negatives, $P( \Em = 0, E =1)$.
However, these two types of errors can bear significantly different
consequences. For instance, incorrectly predicting a hurricane (false
positive) might be inconvenient but acceptable; however, failing to
predict one (false negative) can be catastrophic. To accurately
reflect the distinct impact of false positive and negative, we
introduce the cost-sensitive probability of error:
\begin{equation}
  P_e^{c}( \Qmm,\fm) = c^{+} P( \Em = 1, E =0) + c^{-} P( \Em = 0, E =1),
\end{equation}
where $c^+>0$ and $c^->0$ are the false positive and negative cost
weighting factors, respectively.  \textcolor{black}{These factors
  reflect the relative severity of each type of error, and their
  values are selected according to the specific prediction task. In
  the case of extreme event prediction, the value of $c^-$ is often
  larger than $c^+$. This choice is driven by the understanding that
  non-extreme events occur more frequently than extreme
  ones. Prediction models with equal costs ($c^+=c^-$) are inclined to
  favor the majority class of non-extreme events. By imposing higher
  penalties on false negative errors ($c^->c^+$), we steer the
  prediction model to focus more on accurately identifying the
  critical, but less frequent, extreme events.  The value of the
  factors $c^-$ and $c^+$ is arbitrary, and only their relative
  magnitude matters. Consequently, $c^-$ and $c^+$ can be scaled in
  different manners. To guarantee that the model with the minimum
  probability of error yields $P_e^{c} < 1/2$, we choose $1/c^+ +
  1/c^- = 2$ [see \ref{Appen:A} for more details]}.

\subsection{\textcolor{black}{Minimum cost-sensitive probability of error}}
\label{sec:Pe}

The goal is to estimate the minimum cost-sensitive probability of
error given the observable $\Qmm$ over all possible models $\fm$,
\begin{equation}
    P_{e,\min}^c(\Qmm) = \min_{\fm} P_e^c(\Qmm,\fm).
\end{equation}

\textcolor{black}{The minimum cost-sensitive probability of error
  attainable by any model is [see proof in \ref{Appen:A}]
\begin{equation} 
\label{eq:cs_bayes}
P_{e,\text{min}}^c(\Qmm)
= \mathbb{E}[I(\Qmm)] =  \sum_{\qmm}  I(\Qmm = \qmm)P(\Qmm = \qmm),
\end{equation}
where $\qmm$ is a particular state (i.e., value) for $\Qmm$, $P( \Qmm
= \qmm)$ is the probability of $\Qmm$ taking the value $\qmm$, and
$I(\Qmm = \qmm)$ is the minimum probability of error for the state
$\Qmm = \qmm$:
\begin{equation}
\label{eq:def_I}
I(\Qmm = \qmm) = \min \left\{c^- P(E=1 \mid \Qmm = \qmm),
c^+ \left(1- P(E=1 \mid \Qmm = \qmm ) \right)\right\},
\end{equation}}
where $P(E \mid \Qmm = \qmm)$ is the probability of $E$ conditioned on
$\Qmm = \qmm$.  The minimum error given by Eq.~(\ref{eq:cs_bayes}) is
the consequence of the unavoidable uncertainty intrinsic to chaotic
systems. This uncertainty arises from the lack of knowledge about the
variables \textcolor{black}{(e.g., unobserved variables and/or those
  observed for a limited amount of time)} and errors in the initial
condition values \textcolor{black}{(e.g. finite precision)}, which
transcend the predictive capabilities of any model.

\textcolor{black}{Equation~(\ref{eq:cs_bayes}) provides the precise
  limit for extreme event forecasting; however, its application to the
  development, optimization, and evaluation of models for extreme
  event prediction is challenging due to its non-convex nature. This
  motivates the derivation of information-theoretic lower and upper
  bounds for $P_{e,\min}^c(\Qmm)$ that are more amenable in terms of
  applications and interpretation. For example, obtaining
  $P_{e,\min}^c(\Qmm)$ reliably from Equation~(\ref{eq:cs_bayes}) may
  not be possible in situations where, on the other hand,
  information-theoretic quantities can be efficiently calculated using
  estimators~\cite{paninski2003}. Even when
  Equation~(\ref{eq:cs_bayes}) can be evaluated accurately, its
  manipulation becomes challenging in the context of model development
  due to the non-linearity introduced by the $\min(\cdot)$
  operator~\cite{hastie2009}. In such instances, using an
  information-theoretic formulation of the error facilitates the
  optimization of model parameters. Information theory can also be
  employed for feature selection, specifically identifying the input
  variables that most significantly aid in predicting extreme
  events~\cite{guyon2003}. Additionally, the sources of error
  contributing to $P_{e,\min}^c(\Qmm)$ are more easily interpreted in
  terms of information rather than probabilities, since the former
  adheres to the properties of additivity and the chain
  rule~\cite{cover2006}. In the next section, we derive lower and
  upper bounds for Eq.~(\ref{eq:cs_bayes}) using the framework of
  information theory.}

\subsection{\textcolor{black}{Information-theoretic bounds for minimum probability of error}}
\label{sec:IT_bounds}

The key idea to derive the information-theoretic bounds is that the
prediction of extreme events can be intuitively understood as an
information transmission process, where information from the current
observable state is conveyed to predict the future
state~\citep{LozanoDuran2022}. If the forecast is treated as a noisy
channel, then the Fano's~\citep{Fano1961} and
\textcolor{black}{Hellman's~\citep{hellman1970probability}
  inequalities} provide the foundations for deriving lower and upper
bounds on the minimum probability of error in the transmission of
discrete messages.
%
We measure the uncertainty in the extreme event indicator $E$ given
the information from the observable $\Qmm$ using the cost-sensitive,
conditional \textcolor{black}{Renyi} entropy~\citep{renyi1961measures, Zhao2013}
\begin{equation}
\label{eq:condH_cs}
\begin{aligned}
       H_{\alpha}^c(E \mid \Qmm)
      &= \sum_{\qmm}h_{\alpha}^c\left(P(E=1 \mid \Qmm = \qmm)\right)P(\Qmm = \qmm),
\end{aligned}
\end{equation}
where $$h_{\alpha}^c(p) = h_{\alpha}\left(\min \{c^- p, c^+ (1-p) \}
\right)$$ is the cost-sensitive binary \textcolor{black}{Renyi} entropy
function of order $\alpha> 0$, \textcolor{black}{with $$h_{\alpha}(p) =
  \lim_{\gamma \rightarrow \alpha} \frac{1}{1-\gamma} \log_2
  \left(p^\gamma+(1-p)^\gamma\right).$$}

Equation (\ref{eq:condH_cs}) quantifies the additional information
required to determine the outcome of $E$ given the information in
$\Qmm$ accounting for the weighting factors $c^-$ and $c^+$.
\textcolor{black}{It is useful to interpret $H_{\alpha}^c(E \mid \Qmm)$
  as the uncertainty in $E$ after conducting the `measurement' of
  $\Qmm$. If $E$ and $\Qmm$ are independent random variables, then
  $H_{\alpha}^c(E \mid \Qmm) = H_{\alpha}^c(E)$, i.e., knowing $\Qmm$
  does not reduce the uncertainty in $E$. In this case, $\Qmm$ is not
  a useful observable for forecasting $E$. Conversely, if knowing
  $\Qmm$ provides the knowledge to completely determine $E$, then
  $H_{\alpha}^c(E \mid \Qmm) = 0$, i.e., there is no uncertainty in
  $E$ given $\Qmm$, and the $\Qmm$ can potentially predict $E$ with no
  error. The order $\alpha$ determines the extent to which different
  probabilities influence the uncertainty, with larger values of
  $\alpha$ giving greater weight to higher probabilities.}  For $c^+ =
c^- = \alpha = 1$, $H_{1}^c(E \mid \Qmm)$ is equal to the classic
\textcolor{black}{Shannon} conditional entropy~\citep{Shannon1948},
which is a concave function in the conditional distribution, making it
well-suited for optimization tasks.

\textcolor{black}{The minimum probability of error can be lower and
  upper bounded as a function of the cost-sensitive conditional
  R\'enyi entropy~[see proof in \ref{Appen:A}]
\begin{equation} 
\label{eq:cs_fanos}
 P_{e,\text{min,LB}}^c(\Qmm)
\leq P_{e,\text{min}}^c(\Qmm)
\leq P_{e,\text{min,UB}}^c(\Qmm),
\end{equation}
where the lower and upper bounds are
\begin{equation} 
\label{eq:cs_upper_lower}
\begin{aligned}
 & P_{e,\text{min,LB}}^c(\Qmm) = h_{\alpha}^{-1}\left( H_{\alpha}^c(E \mid \Qmm ) \right) \\
 & P_{e,\text{min,UB}}^c(\Qmm) = 
\min\left\{\frac{1}{2} H_{\alpha}^c(E \mid \Qmm ), C \right\},  \\ 
\end{aligned}
\end{equation}
and $C = \min \left\{c^{-} P(E=1), c^{+}(1-P(E=1))\right\}.$
Eq.~(\ref{eq:cs_fanos}) is valid for $0<\alpha\leq2$, with the
tightest bounds achieved for $\alpha = 2$, i.e., the quadratic
conditional entropy $H_{2}^c(E \mid \Qmm )$. Nonetheless, maintaining
the more general formulation with $\alpha$ is beneficial, as it
establishes a relationship between error and information within the
context of different entropies. 
It is worth noting that the
minimum error in Eq.~(\ref{eq:cs_bayes}) and the information-theoretic
bounds in Eq.~(\ref{eq:cs_fanos}) hold for any value of the
thresholding parameter $\eta$ defining the cutoff for extreme
events}. Furthermore, Eq.~(\ref{eq:cs_bayes}) and
Eq.~(\ref{eq:cs_fanos}) are generally applicable to the prediction of
any binary events, whether they are extreme or not.

\textcolor{black}{A corollary from the conditional entropy inequality
  (a.k.a. \emph{information can't hurt})~\citep{cover2006} is that
  incorporating additional time lags into the vector of observables
  can decrease (but never increase) the minimum probability of
  error~[see proof in \ref{Appen:C}]
  \begin{equation}
    \label{eq:IT_takens}
\begin{aligned}
  & P_{e,\text{min,LB}}^c(\Qmml) \leq P_{e,\text{min,LB}}^c(\Qmmp),
  \ \text{for} \ l > p, \\
  & P_{e,\text{min,UB}}^c(\Qmml) \leq P_{e,\text{min,UB}}^c(\Qmmp),
  \ \text{for} \ l > p, \\
\end{aligned}
\end{equation}  
where $l$ and $p$ denote the number of time lags in $\Qmml$ and
$\Qmmp$, respectively, i.e., $\Qmml = [ \Qm(t), \Qm(t-\delta t_1),
  \dots,\Qm(t-\delta t_l)] \pm \delta \Qmml$ (and similarly for
$\Qmmp$).  The inequality presented in Eq.~(\ref{eq:IT_takens}) is
particularly useful in scenarios where not every variable of the
system is directly observable.  In such instances, it is still
possible to lower the minimum probability of error by employing
additional time-lagged measurements of the observed variables. This
result can be connected to Takens' embedding
theorem~\citep{Takens1981}, whereby the dynamics of a dynamical system
can be captured by embedding a sequence of past observations into a
higher-dimensional space. The latter is consistent with the decrease
in the probability of error from Eq.~(\ref{eq:IT_takens}). Takens'
embedding theorem also states that the delayed-embedding phase space
of $\Qmml$ is topologically equivalent to the original phase space of
the full dynamical system ($\Qf$) for a non-degenerate, noise-free
observable with $l>2 d_A$, where $d_A$ is the dimension of the
attractor. From an information-theoretic viewpoint, this implies that
$\Qmml$ provides the same information as the state vector of the full
system $\Qf$ when there is no uncertainty in the observable ($\delta
\Qmml = \boldsymbol{0}$) and $l>2 d_A$. Under those conditions,
$P_{e,\text{min}}^c(\Qmml) = P_{e,\text{min}}^c(\Qf)=0$ and exact
predictions are possible.}

\textcolor{black}{The minimum probability of error from
  Eq.~(\ref{eq:cs_bayes}) and the information-theoretic bounds from
  Eq.~(\ref{eq:cs_fanos}) are derived assuming the discretization of
  the continuous variable $\Qm$. This reflects the inherent
  uncertainty in the initial conditions, which is captured by the term
  $\delta \Qmm >\boldsymbol{0}$. Nonetheless, the same inequality
  holds when $\Qm$ is assumed to be a continuous variable by replacing
  the $H_{\alpha}^c$ with its continuous extension [see
    \ref{Appen:cont}]. In that case, there is no uncertainty in the
  value of $\Qm$, and we can take $\delta \Qmm = \boldsymbol{0}$. It
  can also be shown that the right-hand side of
  Eq.~(\ref{eq:cs_bayes}) and the bounds in
  Eq.~(\ref{eq:cs_upper_lower}) converge to their continuous
  counterparts when the partition size used to discretize $\Qm$ is
  refined towards zero. Here, we focus on the discrete case, as in
  practical applications there is always some degree of uncertainty in
  $\Qm$. }

\subsection{\textcolor{black}{Sources contributing to the probability of error}}

The \textcolor{black}{probability of error} in the extreme event
forecaster can be decomposed into three sources: $P^c_{e}( \Qmm,\fm) =
P^{c}_{e,I} + P^{c}_{e,O} + P^{c}_{e,M}$ given by
\begin{equation}
\label{eq:error_decomp}
  \begin{aligned}
    P^{c}_{e,I} &= P_{e,\min}^c(\Qfm), \\
    P^{c}_{e,O} &= P_{e,\min}^c(\Qmm) -  P_{e,\min}^c(\Qfm), \\
    P^{c}_{e,M} &= P_{e}^c(\Qmm,\fm) - P_{e,\min}^c(\Qmm),
  \end{aligned}
\end{equation}
where the vector $\Qfm = \Qf \pm \delta \Qf$ contains all the degrees of
freedom governing the system $\Qf$ (i.e., absolute observability) but
with finite precision $\delta \Qf$.  The
  interpretation of each term in Eq.~(\ref{eq:error_decomp}) is as
  follows:
\begin{itemize}
\item[-] $P^{c}_{e,I}$ represents the probability of error solely
arising from uncertainty in the initial conditions. This is because
$\Qfm$ contains all the degrees of freedom of the system, which are
sufficient to integrate the system forward in time. However, the
process is subject to the initial uncertainty $\delta \Qf$ such that
a higher $\delta \Qf$ might result in a higher $P^{c}_{e,I}$. The
magnitude of $\delta \Qf$ varies depending on the specific
problem. If $\Qfm$ is known with infinite precision (i.e.,~$|\delta
  \Qf|=\mathbf{0}$), then $P^{c}_{e,I}$ equals zero.
\item[-] $P^{c}_{e,O}$ denotes the probability of error caused by
missing information from unobserved variables. This error originates
from the fact that $\Qmm$ contains less information than $\Qfm$. As
discussed in~\ref{sec:IT_bounds}, the inclusion of multiple time lags in
$\Qmm$ can compensate for the lack of observed
variables~\cite{Takens1981}. However, $P^{c}_{e,O}$ will still be an
important contributor to the total probability of error in those
situations where the number of degrees of freedom is much larger
than the number of observed variables, $N \gg M$.
\item[-] $P^{c}_{e,M}$ is the probability of error attributable to a
suboptimal model. Values of $P^{c}_{e,M}>0$ imply that $\fm$ is not
efficiently exploiting the information available in $\Qmm$. In those
cases, the model is not operating at its theoretical maximum
performance, and further improvements are 
possible. Conversely, $P^{c}_{e,M}=0$ implies that $\fm$ is the
best-performing model given the observed variables and uncertainties
in the initial conditions.
\end{itemize}

In the following, we demonstrate the application of our results in two
distinct scenarios: the R\"ossler system and the Kolmogorov flow.  The
R\"ossler system offers a simple case for studying extreme events in a
chaotic system where all variables can be observed. We use this case
to illustrate the classification of errors from
Eq.~(\ref{eq:error_decomp}). On the other hand, the Kolmogorov flow,
characterized by complex, multi-scale interactions among numerous
degrees of freedom, represents the dynamics of extreme events found in
more realistic systems. This case is used to demonstrate the effect of
cost-sensitive analysis.

\section{Applications}

\subsection{R\"ossler system}

\begin{figure}[ht]
    \begin{subfigure}[c]{0.49\linewidth}
    \includegraphics[width=\linewidth]{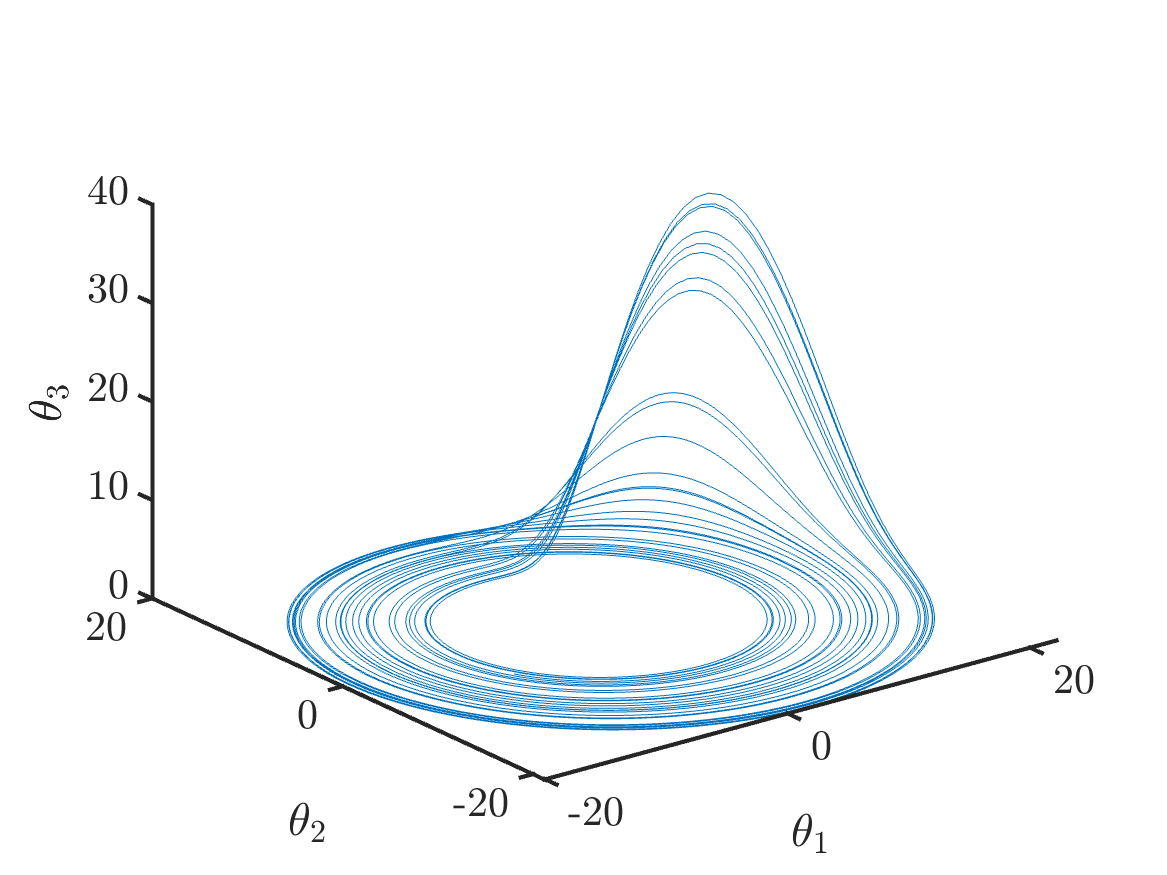}
    \subcaption{}
    \end{subfigure}
    \hspace{3mm}
    \begin{subfigure}[c]{0.49\linewidth}
    \includegraphics[width=\linewidth]{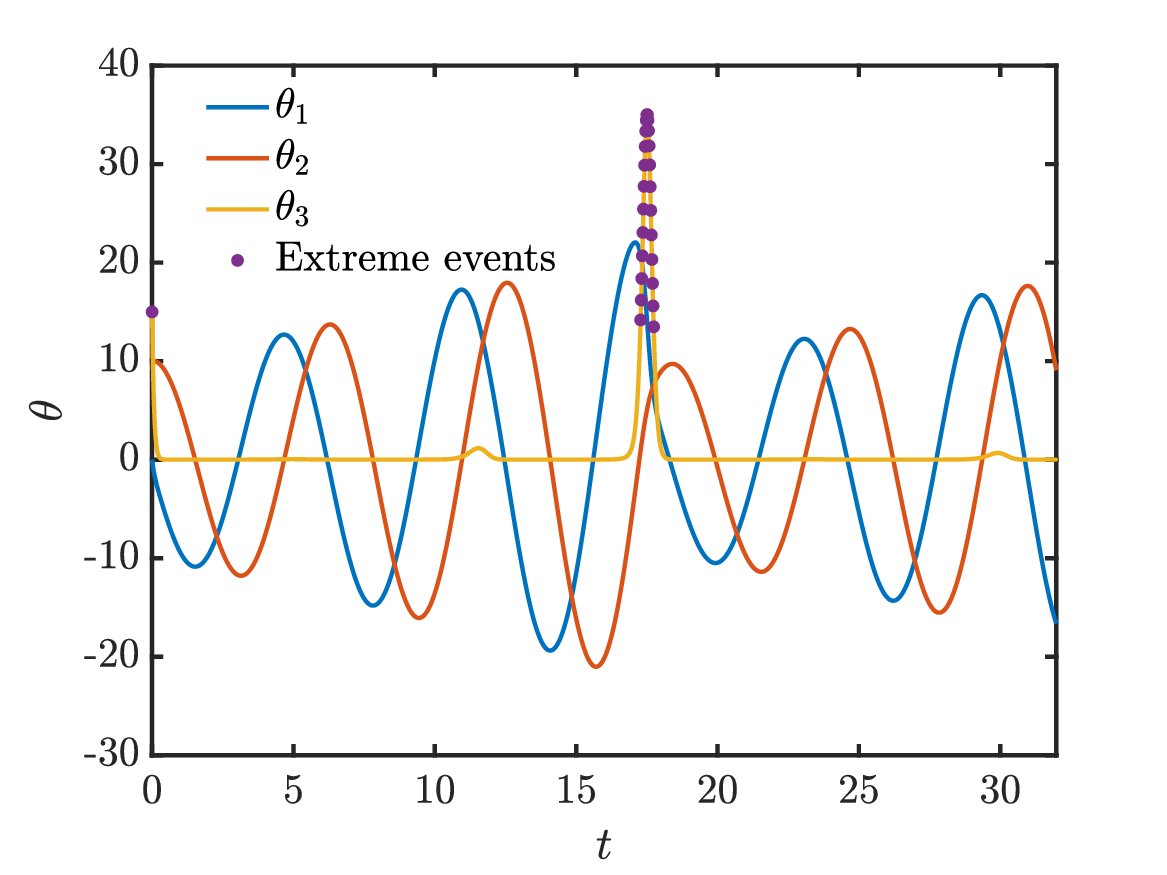}
    \subcaption{}
    \end{subfigure}
    \caption{(a) Trajectory of the R{\"o}ssler system. (b) Extraction
      of time series of $\theta_1$, $\theta_2$, $\theta_3$ and extreme
      events in the R{\"o}ssler system. Although not shown, the whole
      time-span of the signals is 10,000 time
      units.  }
      \label{fig:Rossler_system}
\end{figure}

The R\"ossler system with state variables $\Qf =
[\theta_1,\theta_2,\theta_3]$ is governed by the ordinary differential
equation:
\begin{equation}
\begin{aligned}
  \frac{\dd \theta_1}{\dd t}& = -\theta_2-\theta_3, \\
  \frac{\dd \theta_2}{\dd t}& =  \theta_1+a \theta_2, \\
  \frac{\dd \theta_3}{\dd t}& = b+\theta_3(\theta_1-c),
\end{aligned}
\end{equation}
with parameters $a = 0.1,b = 0.1$, and $c = 14$. We investigate
extreme events in $\theta_3$, which exhibits rare excursions of
intense magnitude.  Figure~\ref{fig:Rossler_system}(a) shows the
trajectory of the R\"ossler system in the three-dimensional phase
space. The extreme event indicator is defined as
\begin{equation}
  E(t)=
  \begin{cases}
    1 & \text { if } \theta_3(t)\geqslant \bar{\theta}_3+ 3\sigma_{\theta_3}, \\
    0 & \text { otherwise, }
  \end{cases}
 \label{eq:Rossler_system} 
\end{equation}
where $\bar{\theta}_3$ and $\sigma_{\theta_3}$ are the mean and
standard deviation of ${\theta_3}$ over time. \textcolor{black}{The
  threshold is set to $\eta = \bar{\theta}_3 + 3\sigma_{\theta_3}$,
  but the conclusions drawn in this section apply to other values of
  $\eta$. Results for a higher threshold can be found in
  \ref{Appen:thres}.}  Figure~\ref{fig:Rossler_system}(b) contains a
fragment of the time history of $\Qf$ and the extreme event indicator
$E$.

\begin{figure}
    \centering
    \begin{subfigure}[c]{0.48\linewidth}
    \includegraphics[width=\linewidth]{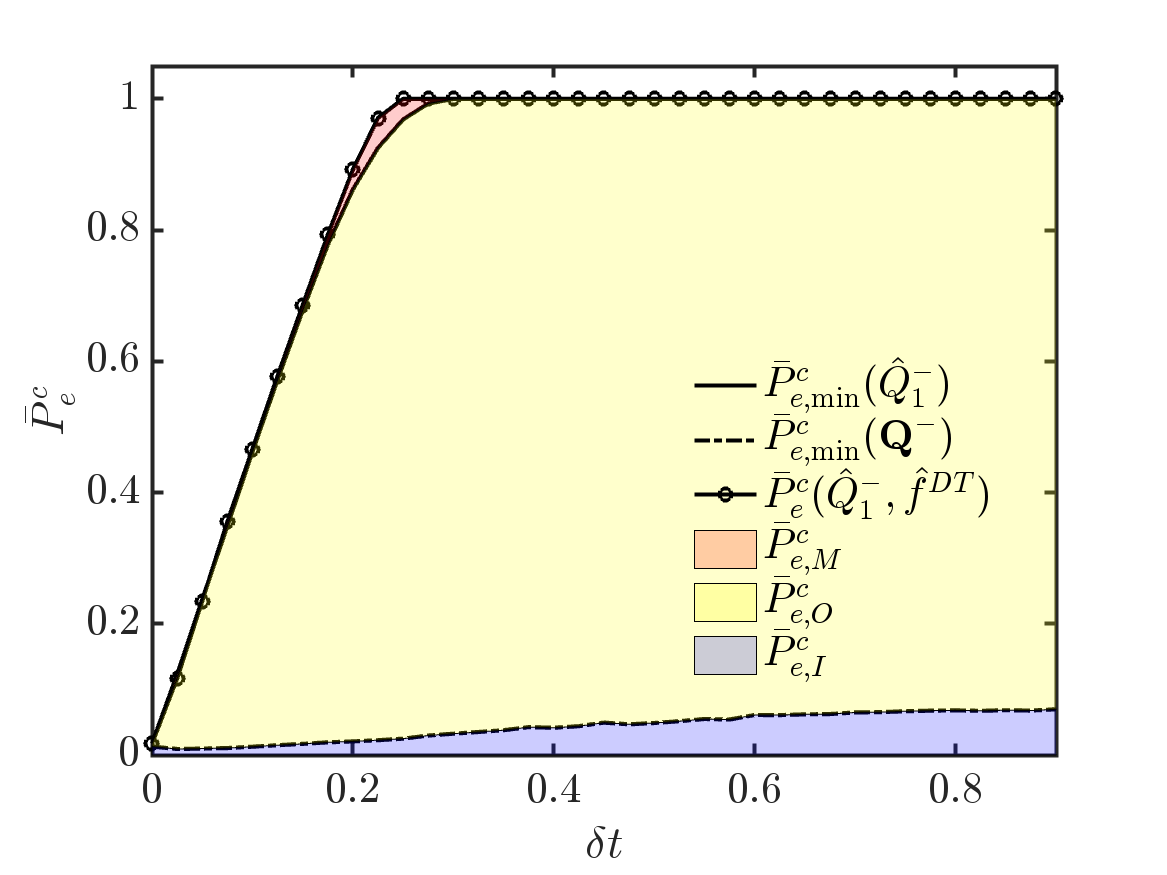}
    \subcaption{}
    \end{subfigure}
    \hspace{1mm}
    \begin{subfigure}[c]{0.48\linewidth}
    \includegraphics[width=\linewidth]{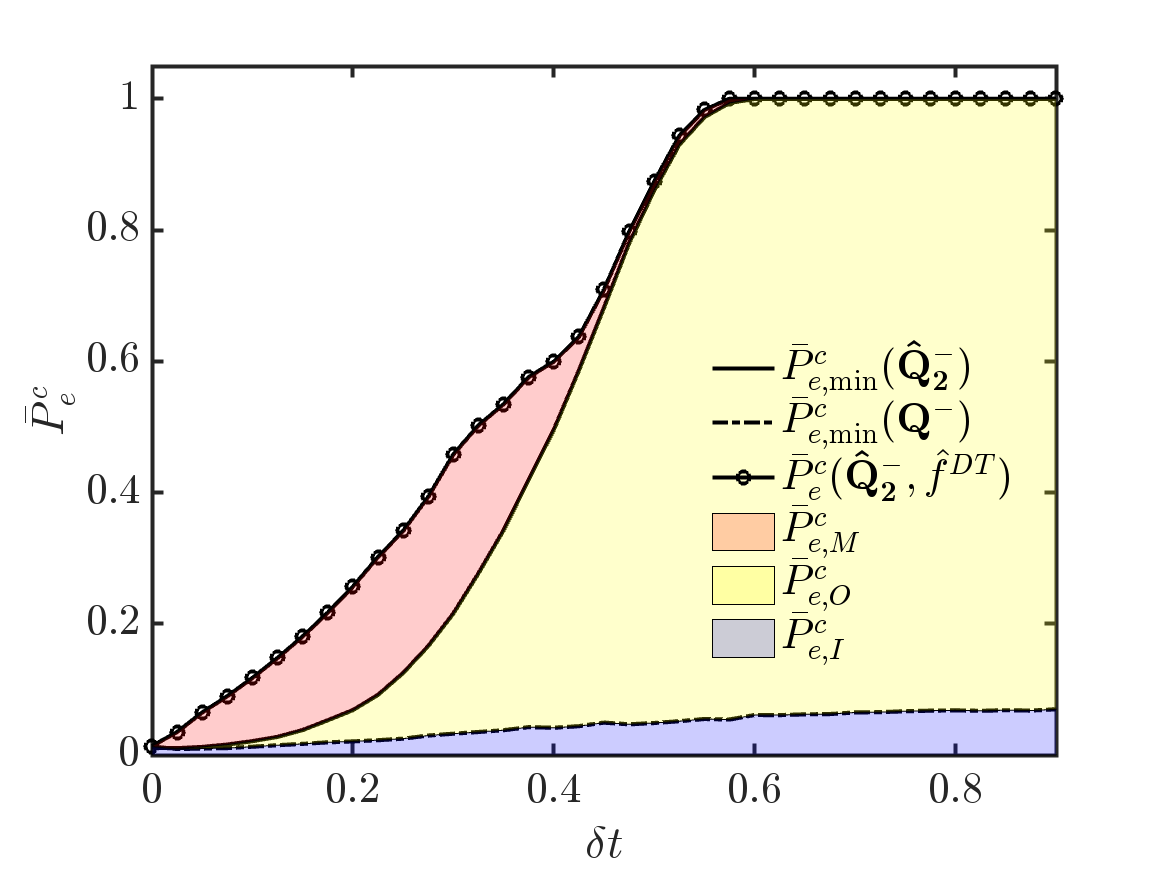}
    \subcaption{}
    \end{subfigure}
    \caption{Normalized probability of error for extreme events
      prediction in the R\"ossler system \textcolor{black}{for the
        threshold $\eta=\bar{\theta}_3+ 3\sigma_{\theta_3}$ using as
        observable (a) $\sQmm_1 = \theta_3(t) \pm \delta \theta_3$ and
        (b) $\Qmm_2= [\theta_3(t), \theta_3(t-\delta
          t),\theta_3(t-2\delta t)]\pm\delta \boldsymbol{\theta}_3$ }.
      $\bar{P}_{e,\min}^c(\sQmm_1)$, $\bar{P}_{e,\min}^c(\Qmm_2)$, and
      $\bar{P}_{e,\min}^c(\Qfm)$ are the minimum probability of error
      using the observable $\sQmm_1$, $\Qmm_2$, and $\Qfm =
      [\theta_1(t),\theta_2(t),\theta_3(t)] \pm \delta
      \boldsymbol{\theta}$, respectively. $\bar{P}^{c}_{e,I}$ (purple)
      is the error due to uncertainty in the initial conditions;
      $\bar{P}^{c}_{e,O}$ (yellow) is the error caused by unobserved
      variables; $\bar{P}^{c}_{e,M}$ (red) is the error due to
      suboptimal model. }
    \label{fig:Rossler_extreme}
\end{figure}

We investigate the case with balanced risk $c^+ = c^- = 1$ and define
the normalized probability of error as $\bar{P}_{e}^c =
P_{e}^c/C$. This normalization is such that $\bar{P}_{e,\min}^c
\rightarrow 1$ for $\delta t \rightarrow \infty$ in practical
applications.  Figure~\ref{fig:Rossler_extreme} shows the normalized
minimum probability of error as a function of time-horizon for extreme
event prediction $\delta t$ \textcolor{black}{using
  Eq.~(\ref{eq:cs_bayes})}. \textcolor{black}{Three} scenarios are
considered.
\begin{itemize}
\item[-] In the first case, we assume that the only observable
  variable is \textcolor{black}{$\sQmm_1 = \theta_3(t) \pm \delta
    \theta_3(t)$}, where the uncertainty in the initial condition is
  set to $\delta \theta_3 = 0.05
  \sigma_{\theta_3}$. \textcolor{black}{Here, the uncertainty $\delta
    \theta_3$ is not introduced by perturbing the equations of the
    system. Instead, the uncertainty is incorporated in a
    non-intrusive manner when calculating the probability
    $P(\theta_3)$ by discretizing $\theta_3$ into bins of size $2\delta
    \theta_3$. This is equivalent to assuming that the solution
    passing through $\theta_3$ cannot be distinguished from another
    trajectory, also contained within the attractor of the system, at
    a distance from $\theta_3$ equal to or less than $\delta
    \theta_3$.}  The associated minimum probability of error,
  $\bar{P}_{e,\min}^c(\sQmm_1)$, is represented by the solid line in
  Fig.~\ref{fig:Rossler_extreme}(a).
\item[-] \textcolor{black}{ In the second scenario, the observable
  includes two time lags in addition to the present time: $\Qmm_2 =
  [\theta_3(t), \theta_3(t-\delta t), \theta_3(t-2\delta t)] \pm
  \delta \boldsymbol{\theta}_3$, where the uncertainty $\delta
  \boldsymbol{\theta}_3$ is again set to $0.05\sigma_{\theta_3}$ for
  all time lags. The minimum probability of error,
  $\bar{P}_{e,\min}^c(\Qmm_2)$, is depicted by the solid line in
  Fig.~\ref{fig:Rossler_extreme}(b). The difference between the two
  minimal errors $\bar{P}_{e,\min}^c(\Qmm_2) $ and
  $\bar{P}_{e,\min}^c(\sQmm_1)$ from Fig.~\ref{fig:Rossler_extreme}(a)
  serves as a measure of the improvement in predictive accuracy gained
  by incorporating observations from two additional times in
  $\theta_3$.}
\item[-] In the \textcolor{black}{third scenario, it is assumed that
  the knowledge of the full state is available at the present time
  with finite precision, i.e., $\Qfm= [\theta_1(t), \theta_2(t),
    \theta_3(t)] \pm \delta \boldsymbol{\theta}$, where $\delta
  \boldsymbol{\theta} = 0.05[\sigma_1,\sigma_2,\sigma_3]$ with
  $\sigma_i$ the standard deviation of $\theta_i$}. The minimum
  probability of error, $\bar{P}_{e,\min}^c(\Qfm)$, is indicated by
  the dashed line in Fig.~\ref{fig:Rossler_extreme}(a) and
  Fig.~\ref{fig:Rossler_extreme}(b).  Errors arising from uncertainty
  in initial conditions are quantified by $\bar{P}^{c}_{e,I} =
  \bar{P}_{e,\min}^c(\Qfm)$ (highlighted by the purple shaded region
  in Fig.~\ref{fig:Rossler_extreme}). The discrepancy between
  \textcolor{black}{$\bar{P}_{e,\min}^c(\sQmm_1)$ and
    $\bar{P}_{e,\min}^c(\Qfm)$} in Fig.~\ref{fig:Rossler_extreme}(a),
  and between \textcolor{black}{$\bar{P}_{e,\min}^c(\Qmm_2)$ and
    $\bar{P}_{e,\min}^c(\Qfm)$} in Fig.~\ref{fig:Rossler_extreme}(b),
  allows us to quantify the errors resulting from the lack of knowledge
  of $\theta_1$ and $\theta_2$ (i.e., $\bar{P}^{c}_{e,O}$, indicated
  by the yellow shaded region in Fig.~\ref{fig:Rossler_extreme}).
\end{itemize}
The region beneath each curve,
\textcolor{black}{$\bar{P}_{e,\min}^c(\sQmm_1)$,
  $\bar{P}_{e,\min}^c(\Qmm_2)$, and $\bar{P}_{e,\min}^c(\Qfm)$},
corresponds to models that are unattainable given the observable
$\sQmm_1$, $\Qmm_2$, and $\Qfm$, respectively, whereas the region
above represents models that are suboptimal.  Over time, all cases
converge to $\bar{P}_e^c \rightarrow 1$ given the chaotic nature of
the system. This convergence is slower for $\bar{P}_{e,\min}^c(\Qfm)$,
as errors are only due to uncertainties in the initial condition.

To illustrate the errors from an actual predictive
  model, we trained decision tree models, $\fm^{DT}$, to predict $E$
using \textcolor{black}{either $\sQmm_1$ or $\Qmm_2$} as
input. \textcolor{black}{Different decision tree models are trained to
  forecast $E$ at each $\delta t$}.  \textcolor{black}{The maximum
  number of branch node splits is 8, and each leaf contains at least
  10 observations.}  The results are also included in
Fig.~\ref{fig:Rossler_extreme}(a) and (b). \textcolor{black}{Additional
  details about the confusion matrix for the decision tree model can
  be found in \ref{Appen:thres}.} The normalized probability of error
for the \textcolor{black}{decision tree} model, \textcolor{black}{Both
  $\bar{P}_{e}^c(\sQmm_1,\fm^{DT})$ and
  $\bar{P}_{e}^c(\Qmm_2,\fm^{DT})$ enable the quantification of the
  model error $\bar{P}^{c}_{e,M}$ for the specific case. The results
  in Fig.~\ref{fig:Rossler_extreme}(a) show that the model error
  $\bar{P}_{e}^c(\sQmm_1,\fm^{DT})$ closely approaches the minimum
  error given by $\bar{P}_{e,\min}^c(\sQmm_1)$, indicating that the
  model is operating near its maximum theoretical performance. On the
  other hand, the results in Fig.~\ref{fig:Rossler_extreme}(b) reveal
  a gap between the model error $\bar{P}_{e}^c(\Qmm_2,\fm^{DT})$ and
  the minimum theoretical error $\bar{P}_{e,\min}^c(\Qmm_2)$ for
  $\delta t<0.4$. This indicates that the model is suboptimal, and
  models with improved performance are possible. For both cases, as
  $\delta t$ increases, $\bar{P}^{c}_{e,O}$ rapidly becomes the
  predominant source of error. Conversely, $\bar{P}^{c}_{e,I}$ is
  minor compared to $\bar{P}^{c}_{e,O}$. Hence, the analysis also
  shows that missing variables have a greater impact on the accuracy
  of the forecast compared to suboptimal modeling and uncertainty in
  the initial conditions. }
\nolinenumbers

\begin{figure}
    \centering
    \includegraphics[width=\tw\linewidth]{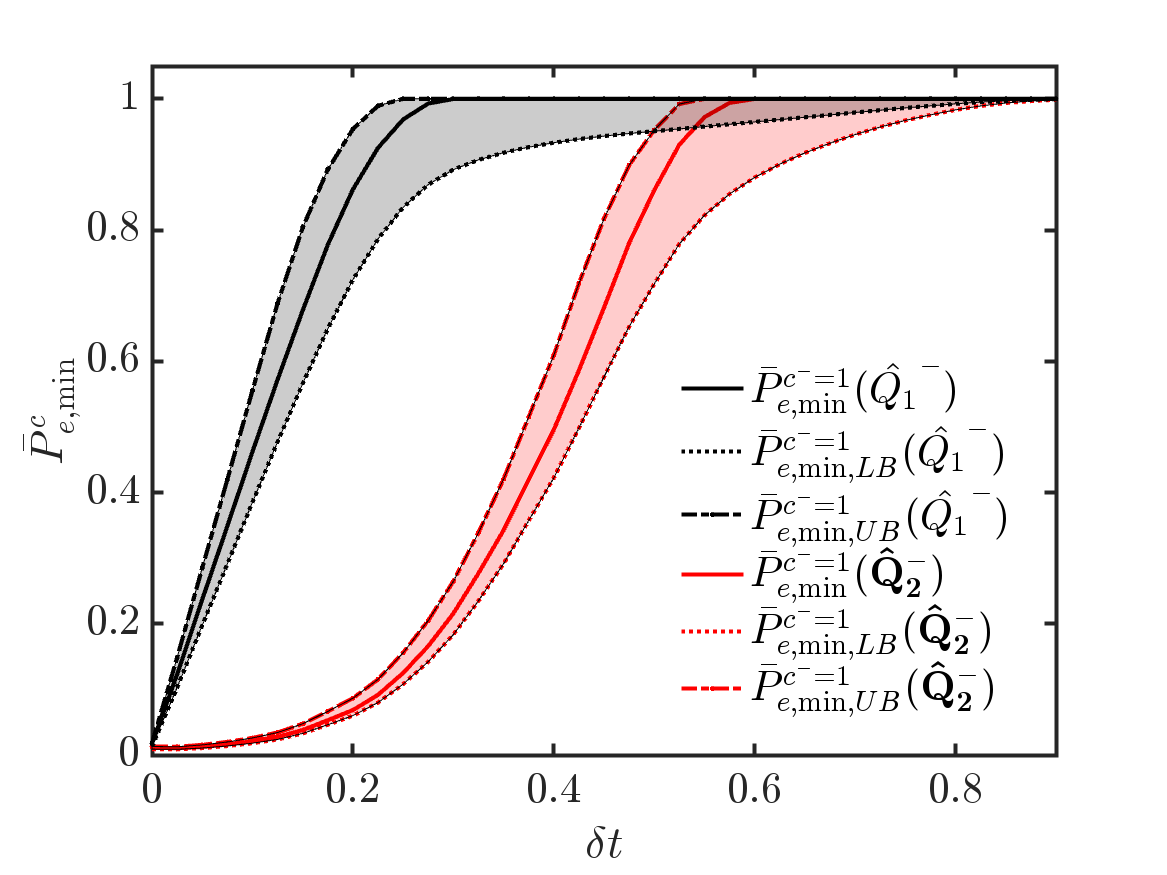}
    \caption{\textcolor{black}{Normalized information-theoretic upper
        and lower bounds of the minimum probability of error for
        extreme events prediction in the Rössler system for the
        threshold $\eta=\bar{\theta}_3 + 3\sigma_{\theta_3}$ and
        observables $\sQmm_1$ and $\Qmm_2$. The solid line is
        $\bar{P}_{e,\min}^c$ and the shaded area represents the region
        confined within $\bar{P}_{e,\text{min,LB}}$ and
        $\bar{P}_{e,\text{min,UB}}$ obtained for the second-order,
        cost-sensitive conditional R\'enyi entropy.}}
    \label{fig:Rossler_extreme_LUB}
\end{figure}

\textcolor{black}{Finally, we compare the exact minimum probability of
  error from Eq.~(\ref{eq:cs_bayes}) with the information-theoretic
  bounds from Eq.~(\ref{eq:cs_upper_lower}) for the second-order,
  cost-sensitive conditional R\'enyi entropy. The results are
  presented in Fig.~(\ref{fig:Rossler_extreme_LUB}) using either
  $\sQmm_1$ or $\Qmm_2$ as observables. In both cases, the bounds
  provide a narrow region within which $\bar{P}_{e,\min}^c$ must be
  confined. In situations where directly obtaining
  $\bar{P}_{e,\min}^c$ is challenging, the region defined by the upper
  and lower bounds can be used to demarcate the theoretical zone of
  near-optimal operation for a model. If the error falls within this
  zone, the model can be considered as possibly operating near its
  best theoretical performance.}

\subsection{Kolmogorov flow}

\begin{figure}[ht]
    \begin{subfigure}[c]{0.49\linewidth}
    \includegraphics[width=\linewidth]{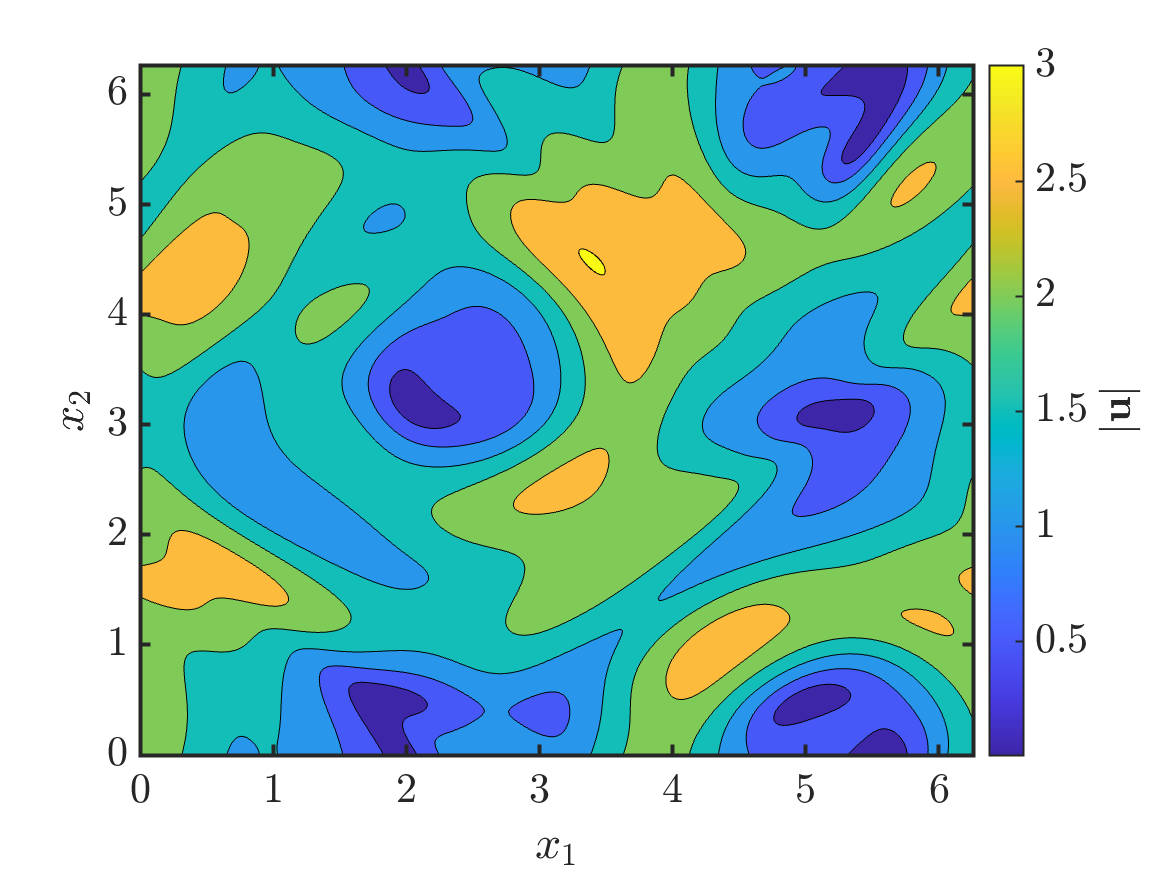}
    \subcaption{}
    \end{subfigure}
    \hspace{3mm}
    \begin{subfigure}[c]{0.49\linewidth}
        \includegraphics[width=\linewidth]{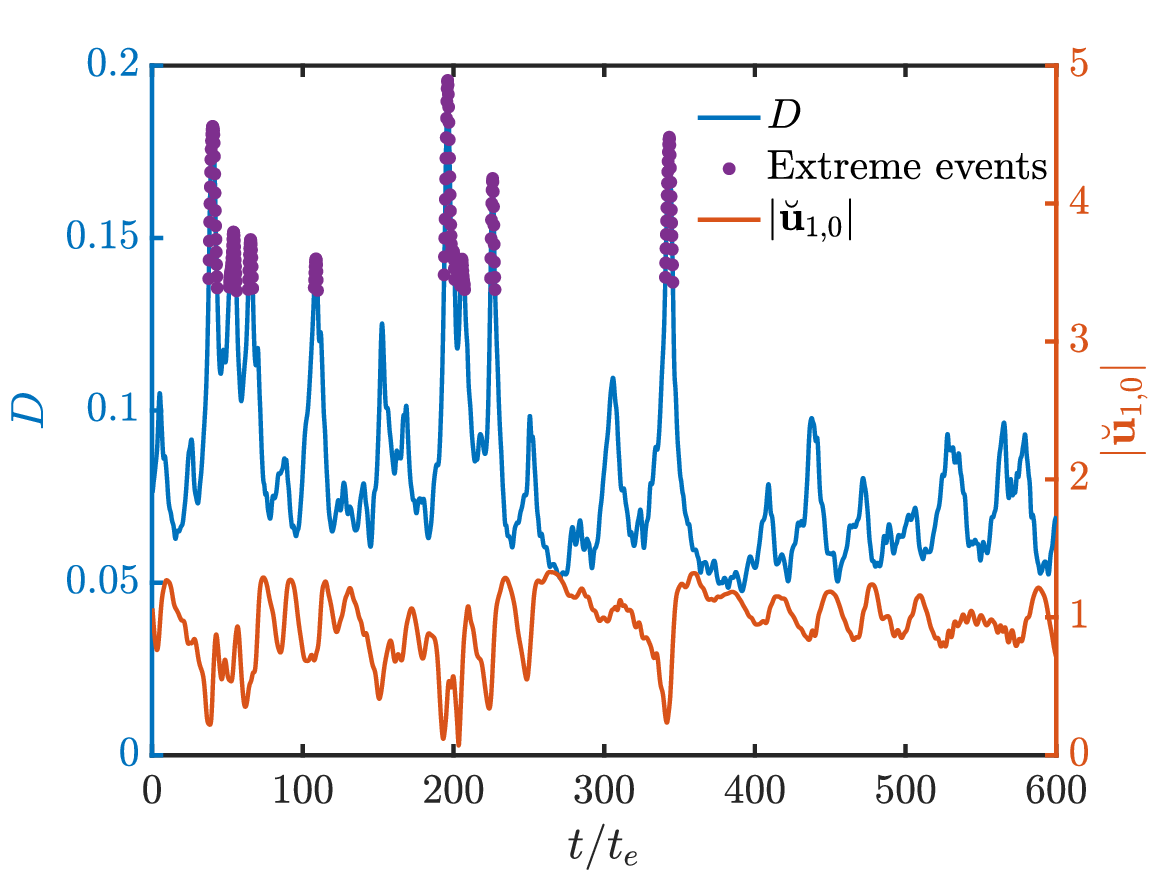}
    \subcaption{}
    \end{subfigure}
    \caption{(a) The instantaneous velocity amplitude $|\boldsymbol
      u|$ for the Kolmogorov flow. (b) Time history of $D(t)$ and
      $|\um_{1,0}|(t)$ \textcolor{black}{for $\eta = \bar{D} + 1.5\sigma_D$.}}
    \label{fig:kol_system}
\end{figure}

Next, we evaluate the cost-sensitive error bounds for forecasting
intense energy dissipation events in a turbulent
flow~\cite{Farazmand2017}.  The case considered is the Kolmogorov
flow: a high-dimensional, chaotic dynamical system described by the
two-dimensional Navier-Stokes equations and driven by monochromatic
body forcing~\cite{Arnold1960}:
\begin{align}
    \frac{\partial u_i}{\partial t} &= -\frac{\partial (u_i u_j) }{\partial x_j} - \frac{\partial \Pi}{\partial x_i}+ \frac{1}{\text{Re}} \frac{\partial^2 u_i}{\partial x_k \partial x_k} + f_i, \quad i=1,2, \\
    \frac{\partial u_i}{\partial x_i} &= 0, 
\end{align}
where repeated indices imply summation, $u_i(x_1,x_2,t)$ is the $i$-th
velocity component, $\Pi$ is the pressure, $x_1$ and $x_2$ are the
spatial coordinates, and $f_i$ is the forcing with $f_1=\sin(4x_2)$
and $f_2=0$. The velocity vector is denoted as $\bu = [u_1, u_2]$.
The flow setup is characterized by the Reynolds number $\text{Re}
= 100$ for which the flow exhibits intermittent bursts of
dissipation events~\cite{Zeff2003}. For our analysis, we use data from
Farazmand and Sapsis \cite{Farazmand2017}, obtained by numerically
resolving all the scales of the problem in a doubly periodic box with
side $2\pi$ and $256^2$ spatial Fourier
modes. Figure~\ref{fig:kol_system}(a) shows the velocity amplitude
$|\boldsymbol u(x_1,x_2,t)|$ at a given time.

Our focus is on the prediction of extreme events characterized by
fluctuations in the mean dissipation rate of kinetic energy $D(t) =
\langle 2 S_{ij} S_{ij}/\text{Re}\rangle$, where $S_{ij} = 1/2
(\partial u_i/\partial x_j + \partial u_j/\partial x_i)$ is the
rate-of-strain tensor, and $\langle \cdot\rangle$ denotes average in
space. The extreme event indicator is
\begin{equation}
\label{eq:kol_extreme}
E(t)= \begin{cases}1 & \text { if } D(t)\geqslant \bar{D}+ 1.5
  \sigma_{D}, \\ 0 & \text { otherwise},\end{cases}
\end{equation}
where $\bar{D}$ and $\sigma_{D}$ are the mean and standard deviation
of ${D}$ over time. \textcolor{black}{Predictions of extreme events
  using a higher threshold can be found in \ref{Appen:C}.}  The
observable chosen is the magnitude of the spatial Fourier mode of
$\bu$ corresponding to the wavenumber $[1,0]$, which is denoted by
$\sQm = \left|\um_{1,0}\right|$. \textcolor{black}{The latter is one of
  the preferred observables for predicting extreme dissipation events
  in the Kolmogorov flow, as it has been shown to correlate with the
  growth of $D$~\cite{Farazmand2017}.}  An excerpt of $\sQm$ and $D$
extracted from the full time history is presented in
Fig.~\ref{fig:kol_system}(b). The time is non-dimensionalized by $t_e
= (\text{Re} \bar{D})^{-1/2}$.
\nolinenumbers

\begin{figure}
  \centering
  \includegraphics[width = \tw\linewidth]{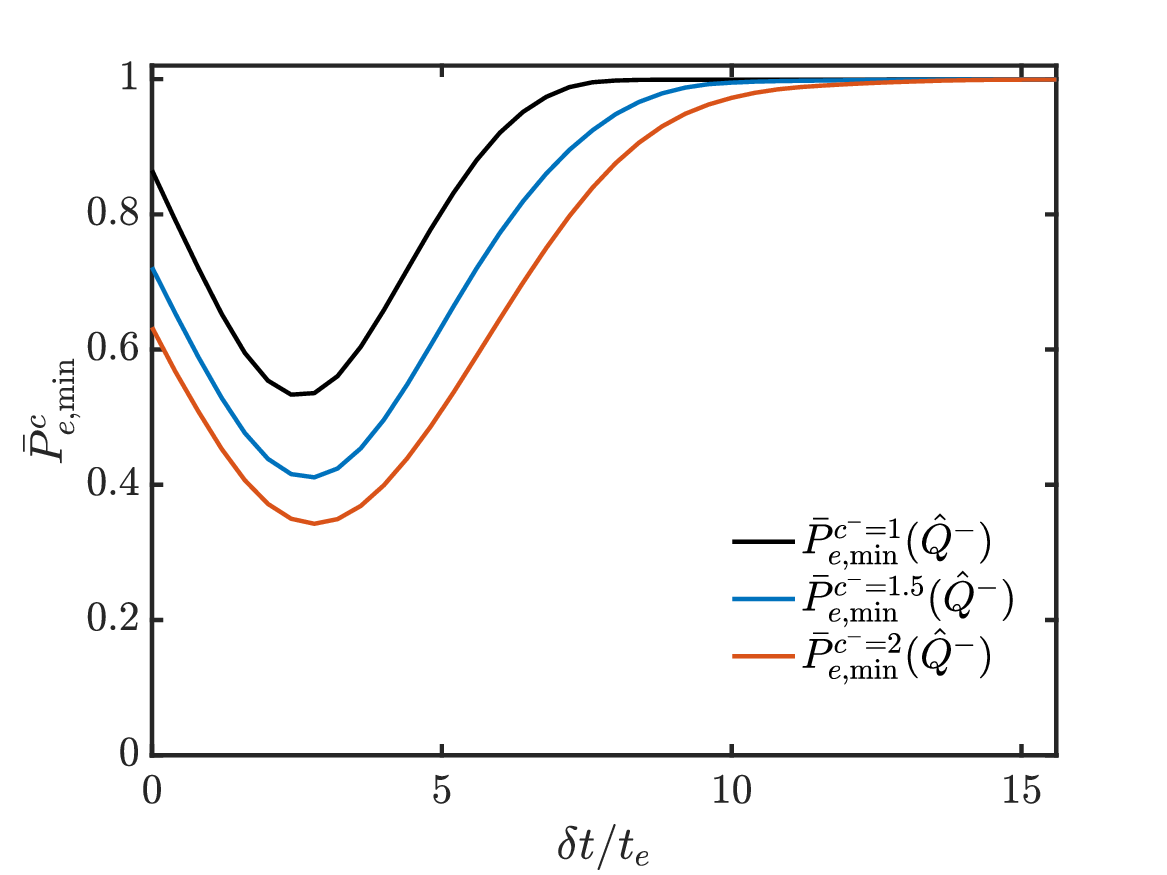}
  \caption{ Normalized, cost-sensitive, minimum probability of error
    for extreme events prediction in the Kolmogorov flow
    \textcolor{black}{for $\eta = \bar{D} + 1.5\sigma_D$} and $c^-=1$,
    1.5, and 2.}
  \label{fig:kol_extreme}
\end{figure}

Three cases are investigated with increasing cost for false positives:
$c^-$ = 1, 1.5, and 2. The first case ($c^-$ = 1) penalizes false
positives and false negatives equally. The other two cases assign a
higher penalty to false negatives, such that the cost of failing to
predict an extreme event is two times (for $c^-=1.5$) or three times
(for $c^-=2$) the cost of incorrectly predicting a non-extreme
event. We denote the minimum probability of error for each case as
$P_{e,\min}^{c^- =1}(\sQmm)$, $P_{e,\min}^{c^- =1.5}(\sQmm)$, and
$P_{e,\min}^{c^- =2}(\sQmm)$,
respectively. Figure~\ref{fig:kol_extreme} shows the results as a
function of $\delta t$.  The minimum cost-sensitive probability of
error is normalized as $\bar{P}_{e,\min}^c = P_{e,\min}^c/C$ such that
$\bar{P}_{e,\min}^c \rightarrow 1$ for $\delta t \rightarrow \infty$.
\nolinenumbers

The results from Fig. \ref{fig:kol_extreme} show that increasing $c^-$
reduces the normalized minimum probability of error, making the
predictions less challenging. This trend is particular to the
Komogorov flow and the chosen variables, and other systems can exhibit
different behavior. Figure~\ref{fig:kol_extreme} also illustrates two
interesting characteristics of $\bar{P}_{e,\min}^{c}(\sQmm)$ for the
three $c^-$ values considered. First, $\bar{P}_{e,\min}^{c}(\sQmm)$ is
not initially zero when $\delta t=0$. This situation arises because
$D$ was not considered an observed variable, leading to uncertainty in
$E$ even at $\delta t=0$. The second interesting observation is that
$\bar{P}_{e,\min}^{c}$ reaches its lowest value at $\delta t_{\min} =
2.8t_e$.  This observation can be understood by noting that in this
system, energy is transferred among different scales due to nonlinear
interactions until it is ultimately dissipated. This process occurs on
a timescale comparable to $\delta t_{\min}$~\citep{Farazmand2017},
which explains the effectiveness of $\left|\um_{1,0}\right|$ in
predicting extreme dissipation events at that time lag.  For times
beyond this point, $\left|\um_{1,0}\right|$ becomes increasingly less
effective due to the chaoticity of turbulence.

\begin{figure}
    \centering
    \includegraphics[width=\tw\linewidth]{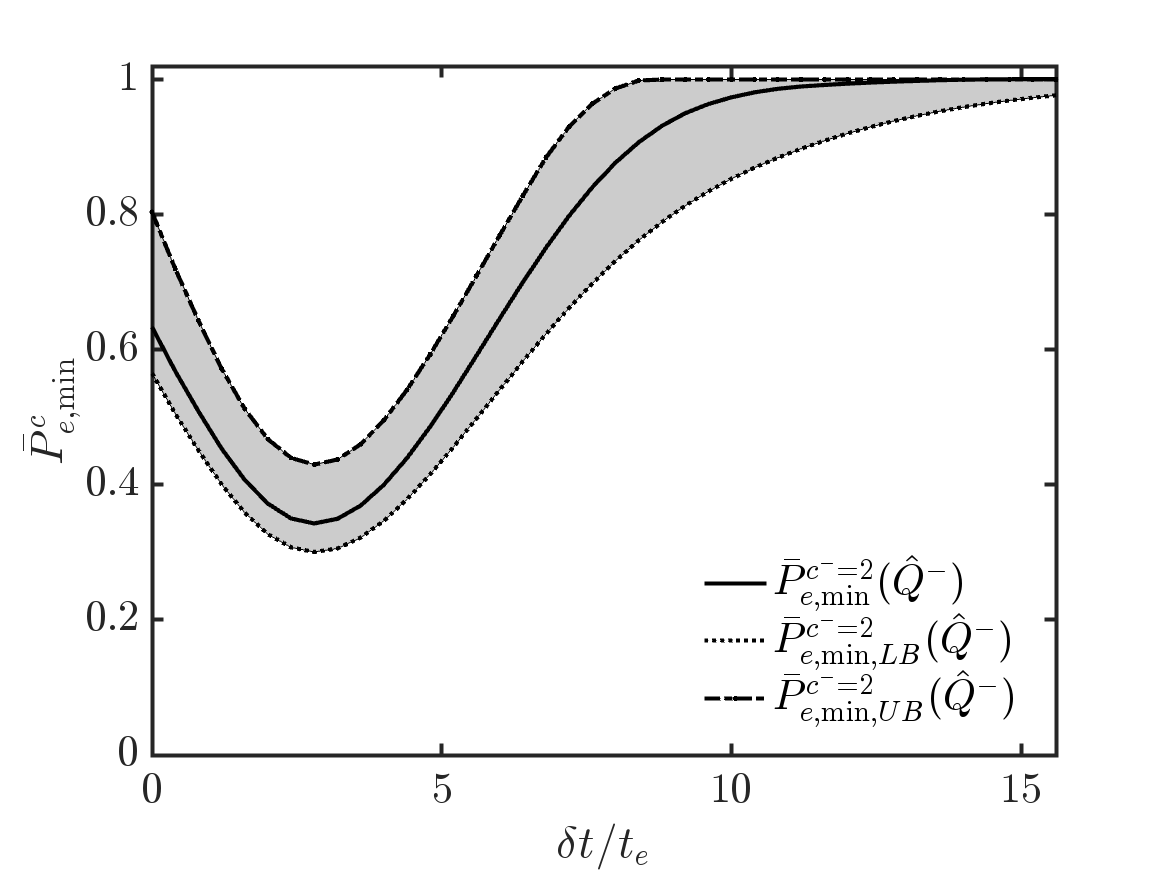}
    \caption{\textcolor{black}{Normalized, cost-sensitive,
        information-theoretic upper and lower bounds of the minimum
        probability of error for extreme events prediction in the
        Kolmogorov flow for $\eta = \bar{D} + 1.5\sigma_D$. The solid
        line is $\bar{P}_{e,\min}^c$ and the shaded area represents
        the region confined within $\bar{P}_{e,\text{min,LB}}$ and
        $\bar{P}_{e,\text{min,UB}}$ obtained for the cost-sensitive,
        second-order, conditional R\'enyi entropy.}}
    \label{fig:kol_extreme_LUB}
\end{figure}

\textcolor{black}{ Finally, we compare the normalized
  information-theoretic upper and lower bounds of the minimum
  probability of error when $c^- = 2$. The results, presented in
  Fig.~\ref{fig:kol_extreme_LUB}, show that the bounds accurately
  reflect the trend observed for $\bar{P}_{e,\min}^{c^-=2}$: there is a
  non-zero minimum probability of error at $\delta t=0$, which
  initially decreases before eventually increasing towards one.}

\section{Conclusions}

In this study, we have derived the minimum cost-sensitive probability
of error in extreme event forecasting (Eq.~\ref{eq:cs_bayes})
\textcolor{black}{along with its information-theoretic lower and upper
  bounds (Eq.~\ref{eq:cs_upper_lower}). The bounds are rooted in the
  cost-sensitive Fano's and Hellman's inequalities for the Rényi
  entropy.} Furthermore, the minimum probability of error and its
bounds are applicable to scenarios involving both balanced and
unbalanced risks. \textcolor{black}{The results are also connected to
  Takens' embedding theorem using the \emph{information can't hurt}
  inequality, which shows that incorporating additional time lags into
  the vector of observables can decrease (but never increase) the
  minimum probability of error.} The probability of error for a
forecasting model was also decomposed into three sources: uncertainty
in the initial conditions, hidden variables, and suboptimal modeling
assumptions (Eq.~\ref{eq:error_decomp}).
 
\textcolor{black}{We have demonstrated the application of these bounds
  to determine the limits of extreme event prediction in two cases
  related to fluid dynamics: the Rössler system and the Kolmogorov
  flow.} These applications illustrate the utility of the minimum
probability of error and its lower and upper bounds as tools for
investigating the intrinsic limitations of forecasting extreme events
in chaotic systems. We have shown that the minimum error and its
bounds maintain their validity irrespective of the chosen modeling
method and play a crucial role in assessing whether models are
functioning near their theoretical optimum.  \textcolor{black}{Future
efforts will be devoted to understanding the limits of predictability
for extreme events such as dissipation, wall-shear stress, and
wall-pressure in turbulent flows at high Reynolds numbers, which are
relevant for advancing the field of external
aerodynamics. Nonetheless, the method presented here is generally
applicable to problems in other fields such as economics, biology, and
finance, among others.}

Finally, the use of Eq.~(\ref{eq:cs_bayes}) and
Eq.~(\ref{eq:cs_upper_lower}) extends beyond the extreme event
predictions presented here; they are broadly applicable to any binary
classification of events, whether they are categorized as extreme or
non-extreme. Eq.~(\ref{eq:cs_bayes}) and Eq.~(\ref{eq:cs_upper_lower})
also lay the foundation for future extensions to forecasting
continuous in time signals by employing generalized versions of Fano's
and \textcolor{black}{Hellman's} inequalities.

\section{Declaration of competing interest}
There is no conflict of interest between the authors and other persons
organizations.

\section{Data availability}
All the data from this research are available upon request.

\section{Acknowledgments}
This work was supported by the National Science Foundation under Grant
No. 2140775 and MISTI Global Seed Funds.

\appendix

\section{Proof of information-theoretic bounds}
\label{Appen:A}
The minimum cost-sensitive probability of error achievable by any
model based on the cost-sensitive uncertainty in $E$ conditioned to
the observable $\Qmm$ is
\begin{equation} 
 P_{e,\text{min}}^c(\Qmm)
= \mathbb{E}[I(\Qmm)],
\end{equation}
and it is lower and upper bounded as
\begin{equation} 
h_{\alpha}^{-1}\left( H_{\alpha}^{c}(E \mid \Qmm ) \right)
\leq P_{e,\text{min}}^c(\Qmm)
\leq \min\left\{\frac{1}{2} H_{\alpha}^c(E \mid \Qmm ), C \right\},
\end{equation}
where \textcolor{black}{$h_\alpha(p)$ is a concave function for $p \in
  [0,1/2]$ when $0<\alpha\leq2$, and $$C = \min \left\{c^{-} P(E=1),
  c^{+}(1-P(E=1))\right\}.$$}
\begin{proof}
The overall cost-sensitive probability of error is a weighted sum of
the probability of error for each specific state of $\Qmm$,
\begin{equation}
\label{eq:condi_decom}
\begin{aligned}
    P_{e}^c(\Qmm,\fm) 
    &= c^{-} P( \Em = 0, E =1) + c^{+} P( \Em = 1, E =0) \\
    &= \sum_{\qmm} \biggl( c^{-} P(\hat{E}=0, E=1 \mid \Qmm = \qmm) \\
    &\qquad + c^{+} P(\hat{E}=1, E=0 \mid \Qmm = \qmm) \biggr) P(\Qmm = \qmm),
\end{aligned}
\end{equation}
where the factors $c^-$ and $c^+$ are scaled as $1/c^- + 1/c^+ = 2$ to
ensure that \textcolor{black}{
\begin{equation}
\begin{aligned}
P^c_{e,\min}(\Qmm) 
 & \leq \begin{cases}
        c^{-} P( E=1 ), 
        & \text{if } \Em  = 0 \\
        c^{+} P( E=0 ), 
        & \text{if } \hat{E}  = 1
    \end{cases} \\
& \leq  \min \left\{c^- P(E=1) ,
 c^+ \left(1- P(E=1 ) \right)\right\}  \\
 & = C  \leq \frac{c^- c^+}{c^+ + c^-} =\frac{1}{2}.
\end{aligned}
\end{equation}
This convention was adopted to avoid $P^c_{e,\min}(\Qmm) > 1/2$, as in
those situations, the model with the minimum probability of error
could be obtained by flipping the model with $P^c_{e,\min}(\Qmm) >
1/2$ to the one with a probability of $1 - P^c_{e,\min}(\Qmm) \leq
1/2$.}
For each specific state of $\Qmm$, the minimum probability of error is determined by 
\begin{equation}
\label{eq:Bayes}
\begin{aligned}
& c^{-} P(\hat{E}=0, E=1 \mid \Qmm = \qmm) +
c^{+} P(\hat{E}=1, E=0 \mid \Qmm = \qmm)\\
& = \begin{cases}
        c^{-} P(\Em=0, E=1 \mid \Qmm = \qmm), & \text{if } \Em(\qmm) = 0 \\
        c^{+} P(\Em=1, E=0 \mid \Qmm = \qmm), & \text{if } \Em(\qmm) = 1
    \end{cases} \\
    & = \begin{cases}
        c^{-} P( E=1 \mid \Qmm = \qmm) - c^{-} \underbrace{P( \Em = 1, E=1 \mid \Qmm = \qmm)}_{0} , 
        & \text{if } \Em(\Qmm= \qmm)  = 0 \\
        c^{+} P( E=0 \mid \Qmm = \qmm) 
        - c^{+} \underbrace{P( \Em =0, E=0 \mid \Qmm = \qmm)}_{0}, 
        & \text{if } \hat{E}(\Qmm=\qmm)  = 1
    \end{cases} \\
&\geq \min \left\{c^- P(E=1 \mid \Qmm = \qmm),
c^+  P(E=0 \mid \Qmm = \qmm)\right\} \\
& = \min \left\{c^- P(E=1 \mid \Qmm = \qmm),
c^+ \left(1- P(E=1 \mid \Qmm = \qmm) \right)\right\}.
\end{aligned}
\end{equation}

Let us define the minimum probability of error at each state as
\begin{equation}
I(\Qmm = \qmm) = \min \left\{c^- P(E=1 \mid \Qmm = \qmm),
c^+ \left(1- P(E=1 \mid \Qmm = \qmm ) \right)\right\}.
\end{equation}
Applying Eq.~(\ref{eq:condi_decom}) and
Eq.~(\ref{eq:Bayes}) we have,
\begin{equation}
\label{eq:Pe_expec}
P_{e,\text{min}}^c(\Qmm)  
=  \sum_{\qmm}  I(\Qmm = \qmm)  P(\Qmm = \qmm) 
 = \mathbb{E}[I(\Qmm)]
 \leq  P_{e}^c(\Qmm,\fm),
\end{equation}
which is the Bayes error rate typically discussed in statistical
classification~\cite{duda1973} but applied here in the context of
extreme event prediction.

\textcolor{black}{Given the concave function $$ h_{\alpha}(p) =
  \lim_{\gamma \rightarrow \alpha}\frac{1}{1-\gamma} \log_2
  \left(p^\gamma+(1-p)^\gamma\right)$$ for $p \in
       [0,0.5]$\cite{ben1978} when $\alpha \in (0,2]$,} the Jensen's
  inequality for the random variable $I(\Qmm)$ results in
\begin{equation}
\label{eq:Jensen's inequality}
h_{\alpha}\left(P_{e,\text{min}}^c(\Qmm)\right)
=h_{\alpha}\left(\mathbb{E}[I(\Qmm)]\right) \geq \mathbb{E}\left[h_{\alpha}\left(I(\Qmm)\right)\right].
\end{equation}
The right hand side of  Eq. (\ref{eq:Jensen's inequality}) is 
\begin{equation}
\begin{aligned}
&\mathbb{E}\left[h_{\alpha}\left(I(\Qmm)\right)\right] 
= \sum_{\qmm} h_{\alpha}\left(I(\Qmm = \qmm)\right) P(\Qmm = \qmm) \\
=& \sum_{\qmm}h_{\alpha}\left(
\min \left\{c^- P(E=1 \mid \Qmm = \qmm),
 c^+ (1- P(E=1 \mid \Qmm = \qmm) )\right\} \right) \\
& P(\Qmm = \qmm) \\
=& \sum_{\qmm}h_{\alpha}^{c}\left(P(E=1 \mid \Qmm = \qmm ) \right)
P(\Qmm = \qmm) =  H_{\alpha}^{c}(E \mid \Qmm), 
\end{aligned}
\end{equation}
where $H_{\alpha}^{c}(E \mid \Qmm)$ is defined as the cost-sensitive
conditional entropy, and the cost-sensitive binary entropy function is
given by
\begin{equation}
h_{\alpha}^{c}(p)= \begin{cases} h_{\alpha}\left(c^- p\right) & \text { for } p \in\left[0, \frac{1}{2 c^-}\right),
\\h_{\alpha} \left(c^+ (1-p) \right) & \text { for } p \in\left[\frac{1}{2 c^-}, 1\right],\end{cases}
\end{equation}
such that
\begin{equation}
\begin{aligned}
h_{\alpha}^{c}(p) = h_{\alpha}\left( \min \left\{c^- p,
c^+ \left(1- p \right)\right\}\right).
\end{aligned}
\end{equation}
\textcolor{black}{ The function $h_{\alpha}^c(p)$ emerges naturally as
  a measure of information. The factors $c^+$ and $c^-$ weight the
  importance (risk) of each event, whereas the order $\alpha$ controls
  how the different probabilities in the distribution contribute to
  the overall measure of uncertainty.  For increasing values of
  $\alpha$, the measure gives more weight to larger probabilities.  For
  example, consider the process of tossing a coin where both heads and
  tails are assigned equal importance ($c^+$ and $c^-$ for heads and
  tails, respectively, with $c^+=c^-=1$). The greatest amount of
  information (i.e., uncertainty) regarding the outcome corresponds to
  the probability $p=1/2$ that maximizes the binary entropy function
  $h_{\alpha}(p)$. However, when the importance of the outcomes
  differs (e.g., heads is preferred over tails, $c^+>c^-$), the
  greatest uncertainty is achieved at $p=\frac{c^+}{c^+ + c^-}$,
  maximizing cost-sensitive binary entropy $h_{\alpha}^c(p)$.}
In conclusion,
\begin{equation}
\label{eq:inequ_con}
    H_{\alpha}^{c}(E \mid \Qmm) 
    = \mathbb{E}\left[h_{\alpha}\left(I(\Qmm)\right)\right] 
    \leq h_{\alpha}\left(\mathbb{E}[I(\Qmm)]\right) 
   =  h_{\alpha}\left(  P_{e,\text{min}}^c(\Qmm)\right),
\end{equation}
which implies
\begin{equation} 
\label{eq:cs_fanos_proof}
h_{\alpha}^{-1}\left( H_{\alpha}^{c}(E \mid \Qmm ) \right)
\leq P_{e,\text{min}}^c(\Qmm)  
= \mathbb{E}\left[I(\Qmm)\right] .
\end{equation}

On the other hand, given the binary entropy function
$h_{\alpha}(p)$ for $p \in [0,0.5], \alpha \in (0,2]$, it is straight
  forward to show that
\begin{equation}
h_{\alpha}\left(I(\Qmm = \qmm)\right) \geq 
2 I(\Qmm = \qmm).
\end{equation}
Applying the expectation operator to each side of the inequality,
\begin{equation}
\label{eq:cs_Hellman}
 H_{\alpha}^{c}(E \mid \Qmm )
=\mathbb{E}\left[h_{\alpha}\left(I(\Qmm)\right) \right]
\geq 2 \mathbb{E}\left[I(\Qmm) \right]
= 2 P_{e,\text{min}}^c(\Qmm).
\end{equation}

In conclusion, 
\begin{equation}
h_{\alpha}^{-1}\left( H_{\alpha}^{c}(E \mid \Qmm ) \right) \leq
P_{e,\text{min}}^c(\Qmm) = \mathbb{E}\left[I(\Qmm) \right] \leq
\min\left\{\frac{1}{2} H_{\alpha}^c(E \mid \Qmm ), C \right\}
\end{equation}
\color{black}
\end{proof}



\section{Proof of inequality of minimum probability of error for additional time lags}
\label{Appen:C}
A consequence of incorporating additional time lags into the vector
  observable variables is that
  \begin{equation}
\begin{aligned}
  & P_{e,\text{min,LB}}^c(\Qmml) \leq P_{e,\text{min,LB}}^c(\Qmmp),
  \ \text{for} \ l > p, \\
  & P_{e,\text{min,UB}}^c(\Qmml) \leq P_{e,\text{min,UB}}^c(\Qmmp),
  \ \text{for} \ l > p, \\
\end{aligned}
\end{equation}  
where $l$ and $p$ denote the number of time lags in $\Qmml$ and
$\Qmmp$, respectively, i.e., 
\begin{equation}
\begin{aligned}
\Qmml &= [ \Qm(t), \Qm(t-\delta t_1),
  \dots,\Qm(t-\delta t_l)] \pm \delta \Qmml,\\
\Qmmp &= [ \Qm(t), \Qm(t-\delta t_1),
  \dots,\Qm(t-\delta t_p)]\pm \delta \Qmmp.
\end{aligned}
\end{equation}
\begin{proof}
First, we prove the cost-sensitive conditional entropy inequality,
\begin{equation}
\label{eq:cost_cond_inequality}
H_{\alpha}^{c}(E \mid \Qmml )  \leq  H_{\alpha}^{c}(E \mid \Qmmp ).
\end{equation}

Noting that $\Qmml  = [\Qmmp, \Qmmr]$, where $\Qmmr = [\Qm(t-\delta t_{p+1}),
  \dots,\Qm(t-\delta t_l)] \pm \delta \Qmmr$, proving Eq~(\ref{eq:cost_cond_inequality}) is equivalent to proving
  \begin{equation}
H_{\alpha}^{c}(E \mid \Qmmp, ~\Qmmr )  \leq  H_{\alpha}^{c}(E \mid \Qmmp ).
\end{equation}

By the law of total probability applied to the conditional probability:
\begin{equation}
\begin{aligned}
\label{eq:prob_decomp}
& P(E=1 \mid \Qmmp = \qmmp)\\
= & \sum_{\qmmr} P(E=1 \mid \Qmmp = \qmmp,~ \Qmmr = \qmmr) P(\Qmmr = \qmmr \mid \Qmmp = \qmmp). \\
\end{aligned}
\end{equation}

The cost-sensitive entropy function $h_{\alpha}^{c}$ is a concave
function when $\alpha \in (0,2]$, applying Jensen's inequality,
\begin{equation}
\begin{aligned}
\label{eq:cs_inequality}
& h_{\alpha}^{c}\left(P(E=1 \mid \Qmmp = \qmmp ) \right) \\
\overset{\ref{eq:prob_decomp}}{=} &  h_{\alpha}^{c}\left(\sum_{\qmmr} P(E=1 \mid \Qmmp = \qmmp,~ \Qmmr = \qmmr) P(\Qmmr = \qmmr \mid \Qmmp = \qmmp) \right) \\
\geq & \sum_{\qmmr}h_{\alpha}^{c}\left( P(E=1 \mid \Qmmp = \qmmp,~ \Qmmr = \qmmr)\right) P(\Qmmr = \qmmr \mid \Qmmp = \qmmp).\\
\end{aligned}
\end{equation}

Applying the inequality to the right hand side of Eq.~(\ref{eq:cost_cond_inequality}), we get
\begin{equation}
    \begin{aligned}
& H_{\alpha}^{c}(E \mid \Qmmp )\\
 = & \sum_{\qmmp}h_{\alpha}^{c}\left(P(E=1 \mid \Qmmp = \qmmp ) \right)
P(\Qmmp = \qmmp) \\
\overset{\ref{eq:cs_inequality}}{\geq}  &\sum_{\qmmp} \sum_{\qmmr} h_{\alpha}^{c}\left( P(E=1 \mid \Qmmp = \qmmp,~ \Qmmr = \qmmr) \right) \\
& P(\Qmmr = \qmmr \mid \Qmmp = \qmmp)  P(\Qmmp = \qmmp)\\
= & \sum_{\qmmp} \sum_{\qmmr}h_{\alpha}^{c}\left( P(E=1 \mid \Qmmp = \qmmp,~ \Qmmr = \qmmr) \right) P(\Qmmr = \qmmr ,~ \Qmmp = \qmmp)\\
= & H_{\alpha}^{c}(E \mid \Qmmp ~,\Qmmr) =H_{\alpha}^{c}(E \mid \Qmml) .
\end{aligned}
\end{equation}

Finally, the conditional entropy inequality is applied to the information theoretic-bounds 
\begin{equation}
\begin{aligned}
&P_{e,\text{min,LB}}^c(\Qmml)= h_{\alpha}^{-1} \left( H_{\alpha}^{c}(E \mid \Qmml) \right), \\
& P_{e,\text{min,LB}}^c(\Qmmp)= h_{\alpha}^{-1} \left( H_{\alpha}^{c}(E \mid \Qmmp) \right), \\
&P_{e,\text{min,UB}}^c(\Qmml)=  \min\left\{\frac{1}{2} H_{\alpha}^c(E \mid \Qmml ), C \right\}, \\
&P_{e,\text{min,UB}}^c(\Qmmp)=  \min\left\{\frac{1}{2} H_{\alpha}^c(E \mid \Qmmp ), C \right\} , \\
\end{aligned}
\end{equation}
by taking into account that $P^c_{e,\min}(\Qmmp)\leq 1/2$, 
\begin{equation}
\begin{aligned}
  & P_{e,\text{min,LB}}^c(\Qmml) \leq P_{e,\text{min,LB}}^c(\Qmmp),
  \ \text{for} \ l > p, \\
  & P_{e,\text{min,UB}}^c(\Qmml) \leq P_{e,\text{min,UB}}^c(\Qmmp),
  \ \text{for} \ l > p, \\
\end{aligned}
\end{equation}
\end{proof}

\section{Continuous extension of information-theoretic bound}
\label{Appen:cont}

We show that the bound for minimum cost-sensitive probability of error
also holds for continuous observables just by replacing the
cost-sensitive entropy $H_{\alpha}^c$ by the
cost-sensitive entropy $\mathcal{H}_{\alpha}^C$ conditioned on a continuous variable.  The minimum
cost-sensitive probability of error achievable by any model with continuous $\Qmm$ is
\begin{equation}
h_{\alpha}^{-1}\left( \mathcal{H_{\alpha}}^{c}(E \mid \Qmm ) \right)
\leq P_{e,\text{min}}^c(\Qmm)
= \mathbb{E}[\mathcal{I}(\Qmm)]
\leq 
\min\left\{\frac{1}{2}\left( \mathcal{H_{\alpha}}^{c}(E \mid \Qmm ) \right), C \right\}.
\end{equation}
The  cost-sensitive conditional entropy is defined as
\begin{equation}
\label{eq:condH_cs_conti}
\begin{aligned}
       \mathcal{H_{\alpha}}^c(E \mid \Qmm)
      &= \int_{\Qmm}h_{\alpha}^c\left(P(E=1 \mid \Qmm = \qmm)\right)\rho_{\Qmm}( \qmm) \mathrm{d} \qmm,
\end{aligned}
\end{equation}
where $\rho_{\Qmm}(\qmm)$ is the probability density function of
$\Qmm$ and the integral is conducted over the support of $\Qmm$.  The
minimum probability of error at each value is defined as
\begin{equation}
\mathcal{I}(\Qmm = \qmm) = \min \left\{c^- P(E=1 \mid \Qmm = \qmm),
c^+ \left(1- P(E=1 \mid \Qmm = \qmm ) \right)\right\}.
\end{equation}
\begin{proof}
Similar to Eq.~(\ref{eq:condi_decom}), the overall cost-sensitive
probability of error is a weighted integral of the probability of
error for each value of $\Qmm$,

\begin{equation}
\label{eq:condi_decom_conti}
\begin{aligned}
 P_{e}^c(\Qmm,\fm)
    &= \int_{\Qmm} \biggl( c^{-} P(\hat{E}=0, E=1 \mid \Qmm = \qmm)  \\
    &\qquad + c^{+} P(\hat{E}=1, E=0 \mid \Qmm =\qmm) \biggr) \rho_{\Qmm}( \qmm) \mathrm{d}\qmm,
\end{aligned}
\end{equation}

Applying Eq.~(\ref{eq:condi_decom_conti}) and Eq.~(\ref{eq:Bayes}) we have,
\begin{equation}
 P_{e,\text{min}}^c(\Qmm) = \int_{\Qmm}  \mathcal{I}(\Qmm=\qmm)  \rho_{\Qmm}(\qmm) \mathrm{d} {\qmm}
 = \mathbb{E}[ \mathcal{I}(\Qmm)].
\end{equation}
Analogous to Eq.~(\ref{eq:inequ_con}),
\begin{equation}
     \mathbb{E}\left[h_{\alpha}\left( \mathcal{I}(\Qmm)\right)\right] 
    \leq h_{\alpha}\left(\mathbb{E}[ \mathcal{I}(\Qmm)]\right) 
    =  h_{\alpha}\left( P_{e,\text{min}}^c(\Qmm)\right),
\end{equation}
where 
\begin{equation}
\begin{aligned}
\mathbb{E}\left[h_{\alpha}\left(I(\Qmm)\right)\right] 
&= \int_{\Qmm} h_{\alpha}\left(I(\qmm)\right) \rho_{\Qmm}(\qmm) \mathrm{d}\qmm \\
&= \int_{\Qmm}h_{\alpha}^{c}\left(P(E=1 \mid \Qmm = \qmm ) \right)
\rho_{\Qmm}( \qmm) \mathrm{d}\qmm \\
&= \mathcal{H_{\alpha}}^{c}(E \mid \Qmm).
\end{aligned}
\end{equation}

On the other hand, similar to Eq.~(\ref{eq:cs_Hellman}),
\begin{equation}
 \mathcal{H_{\alpha}}^{c}(E \mid \Qmm )
=\mathbb{E}\left[h_{\alpha}\left(\mathcal{I}(\Qmm)\right) \right]
\geq 2 \mathbb{E}\left[\mathcal{I}(\Qmm) \right]
= 2 P_{e,\text{min}}^c(\Qmm).
\end{equation}

In conclusion, 
\begin{equation}
h_{\alpha}^{-1}\left( \mathcal{H_{\alpha}}^{c}(E \mid \Qmm ) \right)
\leq P_{e,\text{min}}^c(\Qmm)
= \mathbb{E}\left[\mathcal{I}(\Qmm) \right]
\leq \min\left\{\frac{1}{2}\left( \mathcal{H_{\alpha}}^{c}(E \mid \Qmm ) \right), C \right\}.
\end{equation}
\color{black}

\end{proof}
\newpage
\section{Results for higher thresholds and confusion matrix for $\fm^{DT}$}
\label{Appen:thres}

\begin{figure}
    \centering
    \begin{subfigure}[c]{0.48\linewidth}
    \includegraphics[width=\linewidth]{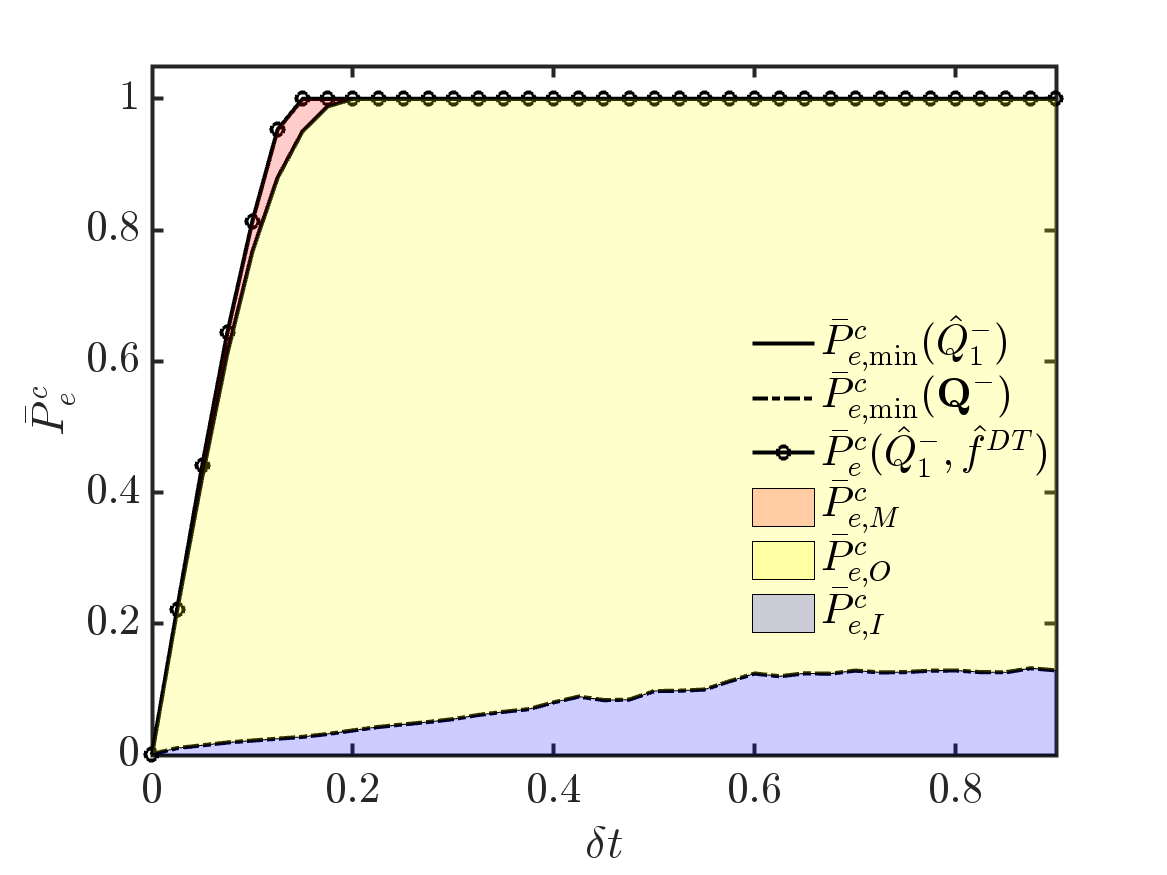}
    \subcaption{}
    \end{subfigure}
    \hspace{1mm}
    \begin{subfigure}[c]{0.48\linewidth}
    \includegraphics[width=\linewidth]{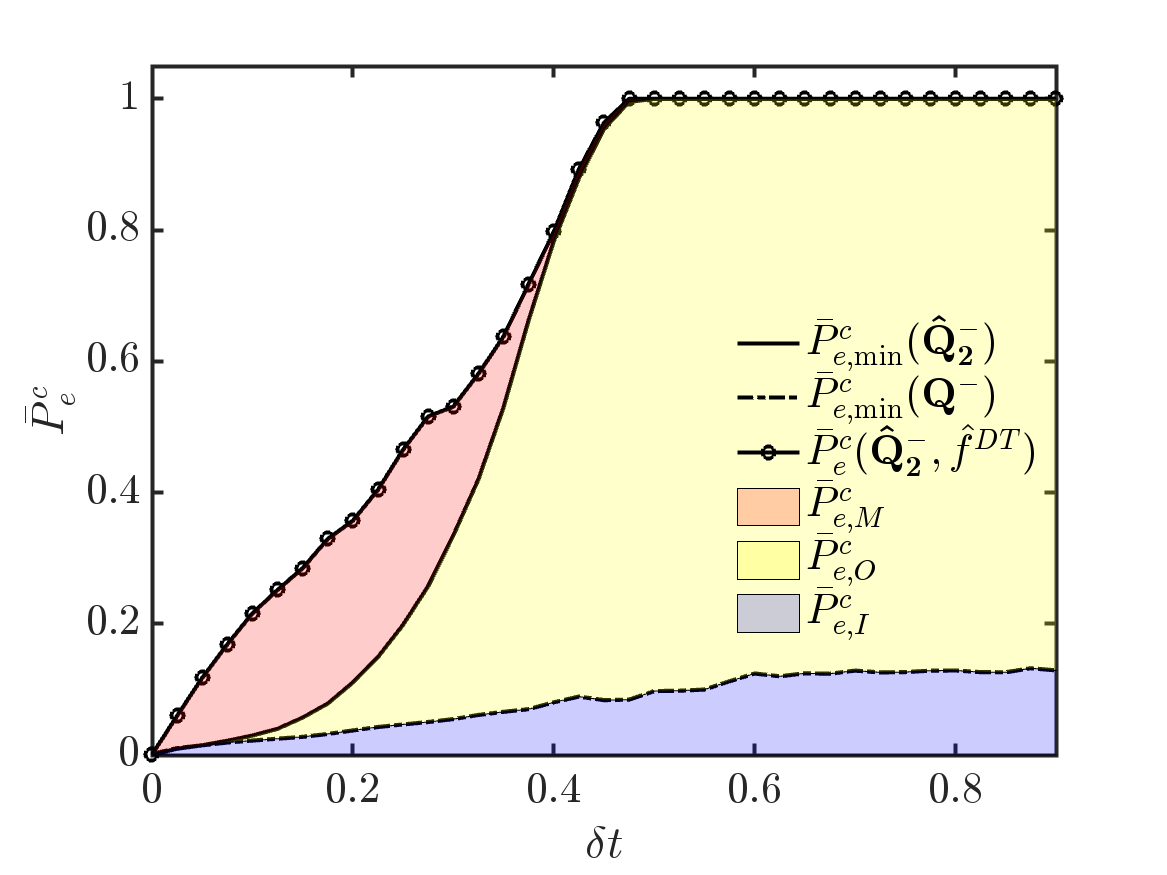}
    \subcaption{}
    \end{subfigure}
    \caption{Normalized probability of error for extreme events
      prediction in the R\"ossler system \textcolor{black}{for the
        threshold $\eta=\bar{\theta}_3+ 6\sigma_{\theta_3}$ using
        observable (a) $\sQmm_1$ and (b) $\Qmm_2$}.
      $\bar{P}_{e,\min}^c(\sQmm_1)$, $\bar{P}_{e,\min}^c(\Qmm_2)$, and
      $\bar{P}_{e,\min}^c(\Qfm)$ are the minimum probability of error
      using the observable $\sQmm_1$, $\Qmm_2$, and $\Qfm$,
      respectively.}
    \label{fig:Rossler_extreme_6}
\end{figure}

\begin{figure}[ht]
  \centering  \includegraphics[width=\tw\linewidth]{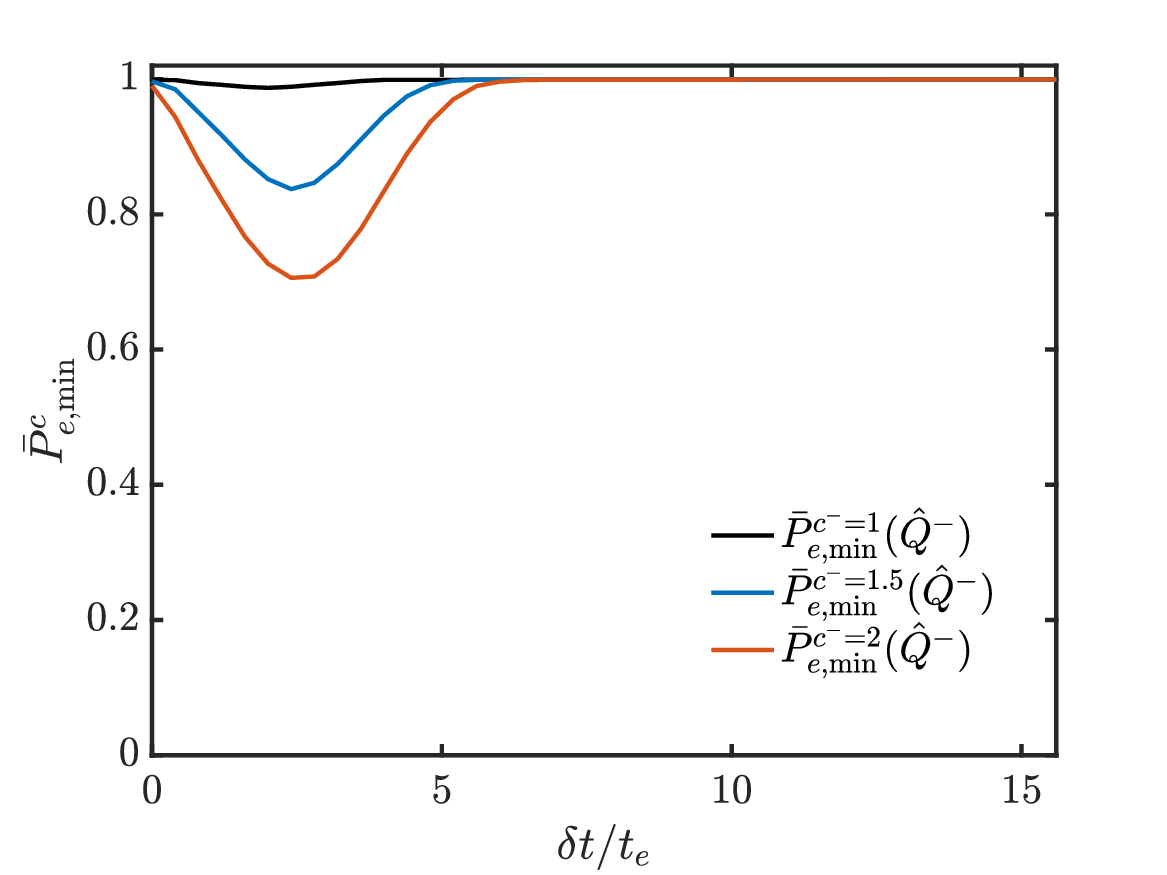}
    \caption{\textcolor{black}{ Normalized,
        cost-sensitive minimum probability of error for extreme event
        prediction in the Kolmogorov flow for $\eta = \bar{D} +
        3\sigma_{D}$.}}
        \label{fig:kol_extreme_3}
\end{figure}

To address the sensitivity of the results to the intensity of the
extreme events in the Rössler system and the Kolmogorov flow, we
repeated the analysis with higher threshold values in both cases. The
minimum probability of error is shown in Fig.~\ref{fig:Rossler_extreme_6} and Fig.~\ref{fig:kol_extreme_3}
as a function of $\delta t$ for the cases discussed above. The main
observation is that increasing the threshold for extreme events makes
the prediction more challenging, due to the higher scarcity of events,
which renders them more unpredictable. Nonetheless, the trends
discussed in the main text regarding the behavior of $\delta t$ remain
unchanged.

\textcolor{black}{ The confusion matrix for the decision tree model to
  predict extreme events in the R\"ossler system using observable $\sQmm_1$ is shown in
  Table~\ref{table:confusion_matrix} for $\delta t = 0.075$.}
 \begin{table}[ht]
    \centering
    \begin{tabular}{cccc}
      \hline
      \textbf{Decision Tree} & \textbf{$\hat{E}=0$} & \textbf{ $\hat{E}=1$} 
      \\ \hline
      \textbf{$E=0$} & 0.9716 & 0.0060 & \\
      \textbf{$E=1$} & 0.0030 & 0.0194 & \\ \hline
    \end{tabular}
    \caption {\textcolor{black}{Confusion Matrix of Decision Tree model. The normalized
      probability of error of the model can be calculated as $\bar
      P_{e}^c(\sQmm_1,\fm^{DT};\delta t = 0.075) = \frac{0.0060 +
        0.0030}{0.0030+0.0194} =
      0.3554$.}}  \label{table:confusion_matrix}
  \end{table}

\color{black}
\bibliographystyle{elsarticle-num} 
\bibliography{references_new}

\end{document}